\documentclass[12pt]{article}
\usepackage{amsmath}
\usepackage{graphicx}
\usepackage{bm}
\usepackage{fancyhdr}
\usepackage{amssymb}
\renewcommand{\theequation}{\arabic{section}.\arabic{equation}}

\begin{document}



\def\a{\alpha}
\def\b{\beta}
\def\d{\delta}
\def\e{\epsilon}
\def\g{\gamma}
\def\h{\mathfrak{h}}
\def\k{\kappa}
\def\l{\lambda}
\def\o{\omega}
\def\p{\wp}
\def\r{\rho}
\def\t{\tau}
\def\s{\sigma}
\def\z{\zeta}
\def\x{\xi}
\def\V={{{\bf\rm{V}}}}
 \def\A{{\cal{A}}}
 \def\B{{\cal{B}}}
 \def\C{{\cal{C}}}
 \def\D{{\cal{D}}}
\def\K{{\cal{K}}}
\def\O{\Omega}
\def\R{\bar{R}}
\def\T{{\cal{T}}}
\def\L{\Lambda}
\def\f{E_{\tau,\eta}(sl_2)}
\def\E{E_{\tau,\eta}(sl_n)}
\def\Zb{\mathbb{Z}}
\def\Cb{\mathbb{C}}

\def\R{\overline{R}}

\def\beq{\begin{equation}}
\def\eeq{\end{equation}}
\def\bea{\begin{eqnarray}}
\def\eea{\end{eqnarray}}
\def\ba{\begin{array}}
\def\ea{\end{array}}
\def\no{\nonumber}
\def\le{\langle}
\def\re{\rangle}
\def\lt{\left}
\def\rt{\right}

\newtheorem{Theorem}{Theorem}
\newtheorem{Definition}{Definition}
\newtheorem{Proposition}{Proposition}
\newtheorem{Lemma}{Lemma}
\newtheorem{Corollary}{Corollary}
\newcommand{\proof}[1]{{\bf Proof. }
        #1\begin{flushright}$\Box$\end{flushright}}

\baselineskip=20pt

\newfont{\elevenmib}{cmmib10 scaled\magstep1}
\newcommand{\preprint}{
   \begin{flushleft}
   \end{flushleft}\vspace{-1.3cm}
   \begin{flushright}\normalsize
   \end{flushright}}
\newcommand{\Title}[1]{{\baselineskip=26pt
   \begin{center} \Large \bf #1 \\ \ \\ \end{center}}}
\newcommand{\Author}{\begin{center}
   \large \bf
Junpeng Cao${}^{a, b}$,Wen-Li Yang${}^{c, d}\,
\footnote{Corresponding author: wlyang@nwu.edu.cn}$,Kangjie
Shi${}^c$ and~Yupeng Wang${}^{a,b} \,\footnote{Corresponding author:
yupeng@iphy.ac.cn}$
 \end{center}}
\newcommand{\Address}{\begin{center}
     ${}^a$Beijing National Laboratory for Condensed Matter
           Physics, Institute of Physics, Chinese Academy of Sciences, Beijing
           100190, China\\
     ${}^b$Collaborative Innovation Center of Quantum Matter, Beijing,
     China\\
     ${}^c$Institute of Modern Physics, Northwest University,
     Xian 710069, China\\
     ${}^d$Beijing Center for Mathematics and Information Interdisciplinary Sciences, Beijing, 100048,  China
   \end{center}}
\newcommand{\Accepted}[1]{\begin{center}
   {\large \sf #1}\\ \vspace{1mm}{\small \sf Accepted for Publication}
   \end{center}}

\preprint \thispagestyle{empty}
\bigskip\bigskip\bigskip

\Title{Nested off-diagonal Bethe ansatz and exact solutions of the
$su(n)$ spin chain with generic integrable boundaries} \Author

\Address \vspace{1cm}

\begin{abstract}
The nested off-diagonal Bethe ansatz method is proposed to
diagonalize multi-component integrable models with generic
integrable boundaries. As an example, the exact solutions of the
$su(n)$-invariant spin chain model with both periodic and
non-diagonal boundaries are derived by constructing the nested $T-Q$
relations based on the operator product identities among the fused
transfer matrices  and the asymptotic behavior of the transfer
matrices.

\vspace{1truecm} \noindent {\it PACS:} 75.10.Pq, 02.30.Ik, 71.10.Pm


\noindent {\it Keywords}: Spin chain; reflection equation; Bethe
Ansatz; $T-Q$ relation
\end{abstract}

\newpage



\section{Introduction}
\label{intro} \setcounter{equation}{0}

The appearance of integrability in planar AdS/CFT \cite{Mal98} is a
rather unexpected occurrence and has led to many remarkable results
\cite{Bei12} (see also references therein) and even ultimately to
the exact solution of planar ${\cal{N}}=4$ supersymmetric Yang-Mills
(SYM) theory. The anomalous dimensions of single-trace operators of
${\cal{N}}=4$ SYM are given by the eigenvalues of certain integrable
closed  spin chain Hamiltonians \cite{Min03,Bei12}. Then it was
shown \cite{Ber05,Hof07} that the computing of the anomalous
dimensions of determinant-like operators of  ${\cal{N}}=4$ SYM can
be mapped to the eigenvalue problem of certain integrable open spin
chain ( spin chain with boundary condition specified by reflection
$K$-matrices or boundary scattering matrices) Hamiltonians
\cite{Mur08,Nep11,Bei12}, while by AdS/CFT the $K$-matrices of the
open chain correspond to open strings attached to maximal giant
gravitons \cite{McG00,Hof07}. Therefore spin chain model has played
an important role in understanding the physical contents of planar
${\cal{N}}=4$ SYM theory and  planar AdS/CFT. Moreover, it has
already provided valuable insight into the important universality
class of boundary quantum physical systems in condensed matter
physics \cite{Duk04}. Motivated by the above great applications, in
this paper, we develop the nested off-diagonal Bethe ansatz method,
a generalization of the method proposed in \cite{cao1}, to solve the
eigenvalue problem of multi-component spin chains with the most
general integrable boundary terms.

So far, there have been several well-known methods for deriving the
Bethe ansatz (BA) solutions of quantum integrable models: the
coordinate BA \cite{Bet31,Alc87,Cra12}, the T-Q approach
\cite{bax1,Bax82,Yan06}, the algebraic BA
\cite{Skl78,Tak79,Kor93,Skl88,Fan96,Bel13}, the analytic BA
\cite{Res83}, the functional BA \cite{Skl92} or the separation of
variables method \cite{Nic12} and many others
\cite{And84,Baz89,Nep04,cao,Yan04-1,Gie05,Gie05-1,Doi06,Baj06,Bas07,Gal08,Fra11,Nie09,Gra10}.
However, there exists a quite usual class of integrable models which
do not possess the  $U(1)$ symmetry (whose transfer matrices contain
not only the diagonal elements but also some off-diagonal elements
of the monodromy matrix and the usual $U(1)$ symmetry is broken,
i.e., the total spin is no longer conserved). Normally, most of the
conventional methods do not work  for these models even though their
integrability has been proven for many years \cite{Skl88}.

Recently, a systematic method  \cite{cao1} for dealing with such
kind of models associated with $su(2)$ algebra  was proposed by the
present authors, which had been shown successfully to construct the
exact solutions of the open Heisenberg spin chain with unparallel
boundary fields, the XXZ spin torus, the closed XYZ chain with odd
site number and other models with general boundary terms
\cite{Li13,Zha13}. With the help of the Hirota equation, Nepomechie
\cite{Nep13-1} generalized the results of \cite{cao1} to the
arbitrary spin XXX open chain with general boundary terms. An
expression for the corresponding eigenvectors  was also proposed
recently in \cite{Bel13-1}.

The central idea of the method in \cite{cao1} is to construct a
proper $T-Q$ ansatz with an extra off-diagonal term (comparing with
the ordinary ones \cite{Bax82}) based on the functional relations
between the transfer matrix (the trace of the monodromy matrix) and
the quantum determinant $\Delta_q(u)$,  at some special points of
the spectral parameter $u=\theta_j$, i.e., \bea
t(\theta_j)t(\theta_j-\eta)\sim\Delta_q(\theta_j). \eea Since the
trace and the determinant are two basic quantities of a matrix which
are independent of the representation basis, this method could
overcome the obstacle of absence of a reference state which is
crucial in most of the conventional Bethe ansatz methods. In this
paper, we propose a nested off-diagonal Bethe ansatz method to solve
the multi-component integrable models (integrable spin chains
associated with  higher rank algebras). This method allows us to
construct the nested $T-Q$ relations based on the recursive operator
product identities and the asymptotic behavior of the transfer
matrices for the systems with both periodic and arbitrary integrable
open boundary conditions. We elucidate our method with the  $su(n)$
spin chain (both periodic and open) model as an example. Our method
might be generalized to the integrable systems associated with
$B_n$, $C_n$ and $D_n$ algebras.

The paper is organized as follows. Section 2 serves as an
introduction of our notations and some basic ingredients. We briefly
describe the inhomogeneous $su(n)$-invariant spin chain with
periodic boundary condition. Based on some operator product
relations for  the antisymmetric fused transfer matrices and their
asymptotic behaviors,  the nested $T-Q$ ansatz of their eigenvalues
and the corresponding Bethe ansatz equations (BAEs) are constructed.
In Section 3, we study the $su(n)$-invariant open spin chains with
general open boundary integrable conditions. Based on some
properties of the $R$-matrix and $K$-matrices, we obtain the
important operator product identities among the fused transfer
matrices of the open chains and their asymptotic behaviors when
$u\longrightarrow \infty$. In section 4, we focus on the
$su(3)$-invariant open spin chain with the most general non-diagonal
boundary terms. The nested Bethe ansatz solution for the eigenvalues
of the transfer matrix and the corresponding Bethe ansatz equations
(BAEs) are given in detail based on the operator product identities
of the transfer matrix and their asymptotic behaviors and values of
the transfer matrices at some special points. The results for the
$su(4)$-invariant spin chain and the $su(n)$-invariant one are given
in Section 5 and Section 6, respectively.  We summarize our results
and give some discussions in Section 7. Some detailed technical
proof is given in Appendix A.

\section{$su(n)$-invariant spin chain with periodic boundary conditions}
\setcounter{equation}{0}

\subsection{Transfer matrix}

Let ${\rm\bf V}$ denote an $n$-dimensional linear space. The
Hamiltonian of $su(n)$-invariant quantum spin system with periodic
boundary condition is given by \cite{3-su375,3-su3751}
\begin{eqnarray}
H=\sum_{j=1}^N P_{j,j+1}, \label{hh}
\end{eqnarray}
where $N$ is the number of sites, $P_{j,j+1}$ is
permutation operator, $P_{ac}^{bd}=\delta_{ad}\delta_{bc}$
with $a, b, c, d=1,\cdots,n$.  The integrability of the system (\ref{hh}) is guaranteed by
the $su(n)$-invariant $R$-matrix $R(u)\in {\rm End}({\rm\bf V}\otimes
{\rm\bf V})$ \cite{3-sun,3-sun1}
\begin{eqnarray}
R_{ij}(u)=\sum_{\alpha,\beta=1}^{n} u e_i^{\alpha \alpha} \otimes
{e}_j^{\beta \beta} + \sum_{\alpha,\beta=1}^{n} \eta e_i^{\alpha
\beta} \otimes {e}_j^{\beta \alpha}, \label{R-matrix}
\end{eqnarray}
where ${e}^{\alpha \beta}$ is the $n\times n$ Weyl matrix with the
definition $(e^{\alpha \beta
})_{\mu\nu}=\delta_{\alpha\mu}\delta_{\beta\nu}$, $\alpha, \beta,
\mu, \nu =1, \cdots, n$, $u$ is the spectral parameter and $\eta$ is
the crossing parameter, respectively. The $R$-matrix can be
expressed in terms of the permutation operator $P$ as
\begin{eqnarray}
&&R_{12}(u)=u+\eta P_{1,2}.\label{R-matrix-1}
\end{eqnarray}
The $R$-matrix satisfies the quantum Yang-Baxter equation (QYBE)
\begin{eqnarray}
 R_{12}(u_1-u_2)R_{13}(u_1-u_3)R_{23}(u_2-u_3)=
 R_{23}(u_2-u_3)R_{13}(u_1-u_3)R_{12}(u_1-u_2), \label{QYB}
\end{eqnarray}
and possesses the following properties:
\begin{eqnarray}
 &&\hspace{-1.45cm}\mbox{
 Initial condition}:\hspace{42.5mm}R_{12}(0)= \eta P_{1,2},\label{Initial}\\
 &&\hspace{-1.5cm}\mbox{
 Unitarity}:\hspace{28.5mm}R_{12}(u)R_{21}(-u)= \rho_1(u)\,{\rm id},\quad \rho_1(u)=-(u+\eta)(u-\eta),\label{Unitarity}\\
 &&\hspace{-1.5cm}\mbox{
 Crossing-unitarity}:\quad
 R^{t_1}_{12}(u)R_{21}^{t_1}(-u-n\eta)
 =\rho_2(u)\,\mbox{id},\quad \rho_2(u)=-u(u+n\eta),
 \label{crosing-unitarity}\\
 &&\hspace{-1.4cm}\mbox{Fusion conditions}:\hspace{22.5mm}\, R_{12}(-\eta)=-2\eta P^{(-)}_{1,2}, \quad
 R_{12}(\eta)=2\eta P^{(+)}_{1,2}.\label{Fusion}
\end{eqnarray}
Here $R_{21}(u)=P_{1,2}R_{12}(u)P_{1,2}$,
$P^{(\mp)}_{1,2}=\frac{1}{2}\{1\mp P_{1,2}\}$ is anti-symmetric
(symmetric) project operator in the tensor product space  ${\rm\bf
V} \otimes {\rm\bf V} $, and $t_i$ denotes the transposition in the
$i$-th space. Here and below we adopt the standard notation: for any
matrix $A\in {\rm End}({\rm\bf V})$, $A_j$ is an embedding operator
in the tensor space ${\rm\bf V}\otimes {\rm\bf V}\otimes\cdots$,
which acts as $A$ on the $j$-th space and as an identity on the
other factor spaces; $R_{ij}(u)$ is an embedding operator of
$R$-matrix in the tensor space, which acts as an identity on the
factor spaces except for the $i$-th and $j$-th ones.

Let us introduce the ``row-to-row"  (or one-row ) monodromy matrix
$T(u)$, which is an $n\times n$ matrix with operator-valued elements
acting on ${\rm\bf V}^{\otimes N}$, \bea T_0(u)
=R_{0N}(u-\theta_N)R_{0\,N-1}(u-\theta_{N-1})\cdots
R_{01}(u-\theta_1).\label{2Mon-V-1} \eea Here
$\{\theta_j|j=1,\cdots,N\}$ are arbitrary free complex parameters
which are usually called as inhomogeneous parameters.

The transfer matrix $t^{(p)}(u)$ of the spin chain with periodic
boundary condition (or closed chain) is given by \cite{Kor93}
\begin{eqnarray}
t^{(p)}(u)=tr_0T_0(u). \label{suntran}
\end{eqnarray}
The QYBE implies that one-row monodromy matrix $T(u)$ satisfies the
following relation
\begin{eqnarray}
R_{00^\prime}(u-v)T_{0}(u)T_{0^\prime}(v)
=T_{0^\prime}(v)T_{0}(u)R_{00^\prime}(u-v).\label{RTT}
\end{eqnarray}
The above equation leads to the fact that the transfer matrices with
different spectral parameters commute with each other:
$[t^{(p)}(u),t^{(p)}(v)]=0$. Then $t^{(p)}(u)$ serves as the
generating functional of the conserved quantities, which ensures the
integrability of the closed spin chain. The  Hamiltonian (\ref{hh})
can be obtained from the transfer matrix as following
\begin{eqnarray}
&&H=\eta \frac{\partial \ln t(u)}{\partial u}|_{u=0,\theta_j=0}.
\end{eqnarray}

\subsection{Operator product identities}
Our main tool is the so-called fusion technique \cite{Kar79}. We shall only consider the
antisymmetric fusion procedure which leads to the desired operator
identities to determine the spectrum of the transfer matrix
$t^{(p)}(u)$ given by (\ref{suntran}).

For this purpose, let us introduce the anti-symmetric projectors
which are determined by the following induction relations \bea
P^{(-)}_{1,2,\cdots,
m+1}=\frac{1}{m+1}\sum\lt(1-P_{1,2}-P_{1,3}-\ldots-P_{1,m+1}\rt)P^{(-)}_{2,3,\cdots,
m+1}, \quad m=1,\ldots,n-1.\no \eea For instance, \bea
&&P^{(-)}_{1,2}=\frac{1}{2}\lt(1-P_{1,2}\rt),\no\\
&&P^{(-)}_{1,2,3}=\frac{1}{6}\lt(1-P_{1,2}-P_{2,3}+P_{1,2}P_{2,3}+P_{2,3}P_{1,2}-P_{1,2}P_{2,3}P_{1,2}\rt).\no
\eea We introduce further the fused one-row monodromy matrices
$T_{\langle 1,\ldots,m\rangle}(u)$ (cf. (\ref{2Mon-V-1})) \bea
 T_{\langle 1,\ldots,m\rangle}(u)=P^{(-)}_{1,2,\ldots, m}\,T_1(u)T_2(u-\eta)\ldots T_m(u-(m-1)\eta)\,P^{(-)}_{1,2,\ldots, m},
 \label{Fused-transfer-period}
\eea and the associated fused transfer matrices $t^{(p)}_m(u)$
\begin{eqnarray}
t^{(p)}_m(u)=tr_{12\cdots m}\{T_{\langle 1,\ldots,m\rangle}(u) \}, \quad m=1, \cdots, n,
\label{tm}
\end{eqnarray} which includes the fundamental transfer matrix $t^{(p)}(u)$
given by (\ref{suntran}) as the first one, i.e.,
$t^{(p)}(u)=t^{(p)}_1(u)$. It follows from the fusion of the
$R$-matrix \cite{Kar79} that the fused transfer matrices constitute
commutative families \bea [t^{(p)}_i(u),\,t^{(p)}_j(v)]=0,\quad
i,j=1,\ldots,n. \eea We note that $t^{(p)}_n(u)$ is the quantum
determinant (proportional to the identity operator for generic $u$
and $\{\theta_j\}$),
\begin{eqnarray}
&& t^{(p)}_n(u)=\Delta^{(p)}_q(u)\times {\rm id}=\prod_{l=1}^N(u-\theta_l+\eta)
\prod_{j=1}^N\prod_{k=1}^{n-1}(u-\theta_j-k\eta)\,\times {\rm id}.\label{Q-det-Period}
\end{eqnarray}
Let us evaluate the product of the fundamental transfer matrix  and the fused ones at some special points
\bea
t^{(p)}(\theta_j)t^{(p)}_{m}(\theta_j-\eta)&=&tr_1\lt\{ T_1(\theta_j)\rt\} tr_{2\cdots m+1}\lt\{T_{\langle 2,\ldots,m+1\rangle}(\theta_j-\eta) \rt\}\no\\
&=&tr_{12\cdots m+1}\lt\{T_1(\theta_j)T_{\langle 2,\ldots,m+1\rangle}(\theta_j-\eta)\rt\} \no\\
&\stackrel{(\ref{A.1})}{=}&tr_{12\cdots m+1}\lt\{P^{(-)}_{1,2,\ldots, m+1}T_1(\theta_j)T_2(\theta_j-\eta)\ldots T_{m+1}(\theta_j-m\eta)
P^{(-)}_{2,\ldots, m+1}\rt\}\no\\
&=&tr_{12\cdots m+1}\lt\{P^{(-)}_{1,2,\ldots, m+1}T_1(\theta_j)T_2(\theta_j-\eta)\ldots T_{m+1}(\theta_j-m\eta)P^{(-)}_{1,2,\ldots, m+1}\rt\}\no\\
&=&tr_{12\cdots m+1}\lt\{T_{\langle 1,\ldots,m+1\rangle}(\theta_j)
\rt\} \eea According to the definition (\ref{tm}), we thus have the
following functional relations among the transfer matrices
\begin{eqnarray}
t^{(p)}(\theta_j)t^{(p)}_m(\theta_j-\eta)=t^{(p)}_{m+1}(\theta_j),\quad m=1,\ldots,n-1,\quad j=1,\cdots,N.
\label{id1}
\end{eqnarray}
The initial condition (\ref{Initial}) and the properties
(\ref{Fusion}) of the $R$-matrix imply that the fused transfer
matrix $t^{(p)}_m(u)$ vanishes at some special points, \bea
 t^{(p)}_m(\theta_j+\eta)=t^{(p)}_m(\theta_j+2\eta)=\ldots=t^{(p)}_m(\theta_j+(m-1)\eta)=0.
\eea This fact allows us to introduce some commutative operators $\{\tau^{(p)}_m(u)\}$ associated with the fused transfer matrices
$\{t^{(p)}_m(u)\}$
\begin{eqnarray}
t^{(p)}_m(u)=\prod_{l=1}^N\prod_{k=1}^{m-1}(u-\theta_l-k\eta)\tau^{(p)}_m(u),\quad [\tau^{(p)}_l(u),\tau^{(p)}_m(v)]=0,
\quad l,m=1,\ldots,n.\label{tau-periodic}
\end{eqnarray}
We use the convention: $\tau^{(p)}(u)=\tau^{(p)}_1(u)$. From the
above equations and the definitions (\ref{tm}) of the fused transfer
matrices, we conclude that the operators $\{\tau^{(p)}_m(u)\}$, as
functions of $u$, are polynomials of degree $N$ with the following
asymptotic behaviors \bea
 \tau_m^{(p)}(u)=\frac{n!}{m!(n-m)!}\,u^N+\ldots,\quad u\rightarrow\infty.\label{Asymptotic-periodic}
\eea
The operator identities (\ref{id1}) implies that these operators satisfy the following functional relations
\begin{eqnarray}
\tau^{(p)}(\theta_j)\tau^{(p)}_m(\theta_j-\eta)=\prod_{l=1}^N(\theta_j-\theta_l-\eta)\tau^{(p)}_{m+1}(\theta_j),\quad
j=1,\ldots,N,\quad m=1,\ldots,n-1.\label{id2}
\end{eqnarray}

\subsection{Nested T-Q relation}

The explicit expression (\ref{Q-det-Period}) of the quantum
determinant, the asymptotic behaviors (\ref{Asymptotic-periodic})
and the functional relations (\ref{id2}) allow one to determine the
eigenvalues of all the operators $\{\tau_m^{(p)}(u)\}$ and
consequently those of $\{t_m^{(p)}(u)\}$  completely with the help
of the relation (\ref{tau-periodic}) as follows. The commutativity
of the transfer matrices with different spectral parameters implies
that they have common eigenstates. Let $|\Psi\rangle$ be a common
eigenstate of $\{t^{(p)}_m(u)\}$, which does not depend upon $u$,
with the eigenvalue $\Lambda^{(p)}_m(u)$, i.e., \bea
t^{(p)}_m(u)|\Psi\rangle=\Lambda^{(p)}_m(u)|\Psi\rangle,\quad
m=1,\ldots n.\no \eea  The analyticity of the $R$-matrix  implies
that the eigenvalues $\Lambda^{(p)}_m(u)$  are polynomials of $u$
with a degree of $mN$. The relations
(\ref{tau-periodic})--(\ref{id2}) give rise to some similar
relations of $\{\Lambda^{(p)}_m(u)\}$  which allow us to determine
$\{\Lambda^{(p)}_m(u)\}$ completely. Here we give the final result.
The proof can be obtained by simple checking the solution satisfying
the resulting relations.

Let us introduce $n$ functions $\{z^{(l)}_p(u)|l=1,\ldots,n\}$,
\begin{eqnarray}
z_p^{(l)}(u) =Q_p^{(0)}(u)
\frac{Q_p^{(l-1)}(u+\eta)Q_p^{(l)}(u-\eta)}{Q_p^{(l-1)}(u)Q_p^{(l)}(u)},\quad l=1,\ldots n,
\label{z}
\end{eqnarray} where the functions $Q_p^{(l)}(u)$  are given by
\begin{eqnarray}
&& Q_p^{(0)}(u)=\prod_{j=1}^N(u-\theta_j), \\
&& Q_p^{(r)}(u)=\prod_{l=1}^{L_r}(u-\lambda^{(r)}_l), \quad r=1,
\ldots, n-1,\\
&& Q_p^{(n)}(u)=1,\label{sun31}
\end{eqnarray} where $\{L_r|r=1,\ldots n-1\}$ are some non-negative integers and the parameters $\{\lambda^{(r)}_l|l=1,\ldots L_r,\,r=1,\ldots n-1\}$ will be
determined by the Bethe ansatz equations (\ref{sunbea221}) (see below).
The eigenvalues $\Lambda^{(p)}_m(u)$ of the $m$-th fused transfer matrix $t^{(p)}_m(u)$ is then given by
\begin{eqnarray}
\Lambda^{(p)}_m(u)=\sum_{1\leq i_1<i_2<\ldots < i_m \leq n}
z_p^{(i_1)}(u)z_p^{(i_2)}(u-\eta) \ldots z_p^{(i_m)}(u-(m-1)\eta),\quad m=1,\ldots,n.
\label{anatz1-2}
\end{eqnarray}
For an example, the eigenvalue $\Lambda^{(p)}(u)$ of the fundamental transfer matrix $t^{(p)}(u)$ is
\begin{eqnarray}
\Lambda^{(p)}(u)&=& Q_p^{(0)}(u+\eta)\frac{Q_p^{(1)}(u-\eta)}{Q_p^{(1)}(u)}
+Q_p^{(0)}(u)\frac{Q_p^{(1)}(u+\eta)Q_p^{(2)}(u-\eta)}{Q_p^{(1)}(u)Q_p^{(2)}(u)}+\ldots\no\\
&&+Q_p^{(0)}(u)\frac{Q_p^{(n-2)}(u+\eta)Q_p^{(n-1)}(u-\eta)}{Q_p^{(n-2)}(u)Q_p^{(n-1)}(u)}+
Q_p^{(0)}(u)\frac{Q_p^{(n-1)}(u+\eta)}{Q_p^{(n-1)}(u)},
\label{anatz1-1-1}
\end{eqnarray} while the eigenvalue $\Lambda^{(p)}_n(u)$ of the fused transfer matrix $t^{(p)}_n(u)$ is
\bea
 \Lambda^{(p)}_n(u)=Q_p^{(0)}(u+\eta)\prod_{l=1}^{n-1}Q_p^{(0)}(u-l\eta),
\eea   which is exactly the quantum determinant $\Delta^{(p)}_q(u)$
given in (\ref{Q-det-Period}). The regular property of
$\Lambda^{(p)}(u)$ implies that the residues of $\Lambda^{(p)}(u)$
at each apparent simple pole $\l^{(r)}_l$ have to vanish. This leads
to the associated BAEs,
\begin{eqnarray}
&&\prod _{j=1,\neq l}^{L_r} \frac
{\lambda_{l}^{(r)}-\lambda_j^{(r)}-\eta}
{\lambda_{l}^{(r)}-\lambda_j^{(r)}+\eta} = \prod_{k=1}^{L_{r-1}}
\frac {\lambda_{l}^{(r)}-\lambda_k^{(r-1)}}
{\lambda_{l}^{(r)}-\lambda_k^{(r-1)}+\eta} \prod_{m=1}^{L_{r+1}}
\frac {\lambda_{l}^{(r)}-\lambda_m^{(r+1)}-\eta}
{\lambda_{l}^{(r)}-\lambda_m^{(r+1)}}, \label{sunbea221}\\
&& \qquad \qquad l=1, \ldots L_{r},\quad r=1,2,\ldots,n-1, \quad L_0=N, \quad L_N=0,
\quad  \lambda_{l}^{(0)}=\theta_l.\no
\end{eqnarray} Moreover, the above Bethe ansatz equations also ensure that the regularities of
all the eigenvalues $\Lambda^{(p)}_m(u)$ given in  (\ref{anatz1-2}), namely, the residues for all
 $\Lambda^{(p)}_m(u)$ at point $\l^{(r)}_l$ vanish. Therefore, the BAEs obtained from all the fused transfer matrices are
self-consistent. Redefine  new parameters  $\bar{\lambda}_j^{(r)} =
\lambda_j^{(r)} -r\eta/2$, the resulting BAEs recover those obtained
by other BAs \cite{3-sun,3-sun1}.
\begin{eqnarray}
&&\prod _{j=1,\neq l}^{L_r} \frac
{\bar{\lambda}_{l}^{(r)}-\bar{\lambda}_j^{(r)}-\eta}
{\bar{\lambda}_{l}^{(r)}-\bar{\lambda}_j^{(r)}+\eta} = \prod_{k=1}^{L_{r-1}}
\frac {\bar{\lambda}_{l}^{(r)}-\bar{\lambda}_k^{(r-1)}-\eta/2}
{\bar{\lambda}_{l}^{(r)}-\bar{\lambda}_k^{(r-1)}+\eta/2} \prod_{m=1}^{L_{r+1}}
\frac {\bar{\lambda}_{l}^{(r)}-\bar{\lambda}_m^{(r+1)}-\eta/2}
{\bar{\lambda}_{l}^{(r)}-\bar{\lambda}_m^{(r+1)}+\eta/2}, \label{sunbea2210}\\
&& \qquad \qquad l=1, \ldots L_{r},\quad r=1,2,\ldots,n-1, \quad L_0=N, \quad L_N=0,
\quad  \lambda_{l}^{(0)}=\theta_l.\no
\end{eqnarray}
Finally, we take the homogeneous limit $\theta_j\to 0$. In this
case, the eigenvalue of the Hamiltonian (\ref{hh}) can be expressed
in terms of the Bethe roots
\begin{eqnarray}
E= \sum_{l=1}^{L_1}
\frac{\eta^2}{(\bar{\lambda}_l^{(1)}+\frac\eta2)(\bar{\lambda}_l^{(1)}-\frac\eta2)}+N.
\end{eqnarray}

\section{$su(n)$-invariant spin chain with general open boundary conditions}
\setcounter{equation}{0}

\subsection{Transfer matrix}

Integrable open chain can be constructed as follows
\cite{Alc87,Skl88}. Let us introduce a pair of $K$-matrices $K^-(u)$
and $K^+(u)$. The former satisfies the reflection equation (RE)
\bea &&R_{12}(u_1-u_2)K^-_1(u_1)R_{21}(u_1+u_2)K^-_2(u_2)\no\\
 &&~~=
K^-_2(u_2)R_{12}(u_1+u_2)K^-_1(u_1)R_{21}(u_1-u_2),\label{RE-V}
\eea and the latter  satisfies the dual RE
\bea
&&R_{12}(u_2-u_1)K^+_1(u_1)R_{21}(-u_1-u_2-n\eta)K^+_2(u_2)\no\\
&&~~~~~~= K^+_2(u_2)R_{12}(-u_1-u_2-n\eta)K^+_1(u_1)R_{21}(u_2-u_1).
\label{DRE-V}\eea
For open spin-chains, instead of the standard
``row-to-row" monodromy matrix $T(u)$ (\ref{2Mon-V-1}), one needs to
consider  the
 ``double-row" monodromy matrix ${\cal{J}}(u)$
\bea
  {\cal{J}}_0(u)&=&T_0(u)K_0^-(u)\hat{T}_0(u),  \label{Mon-V-0}\\
  \hat{T}_0(u)&=&R_{01}(u+\theta_1)R_{02}(u+\theta_{2})\ldots
  R_{0N}(u+\theta_N).\label{Mon-V-2}
\eea Then the double-row transfer matrix $t(u)$ of the open spin
chain is given by \bea
t(u)=tr_0\{K^+_0(u){\cal{J}}_0(u)\}.\label{trans} \eea From the QYBE
and the (dual) RE, one may check that the transfer matrices with
different spectral parameters commute with each other:
$[t(u),t(v)]=0$. Thus $t(u)$ serves as the generating functional of
the conserved quantities, which ensures the integrability of the
system.

In this paper, we consider a generic solution $K^{-}(u)$ to the RE associated with the
$R$-matrix (\ref{R-matrix-1}) \cite{fra,mar,de1,de2,de3}
\begin{eqnarray}
K^-(u)=\xi + u M, \quad M^2=1,\label{km-}
\end{eqnarray}
where $\xi$ is a boundary parameter and $M$ is an $n\times n$
constant matrix (only depends on boundary parameters). Besides the
RE, the $K$-matrix satisfies  the following properties \bea
  K^-(0)=\xi,\quad K^-(u)=u\,M+\ldots,\quad u\rightarrow\infty.\label{K-properties}
\eea Since the second power of $M$ becomes the $n\times n$ identity
matrix, the eigenvalues of $M$ must be $\pm 1$. Suppose that there
are $p$ positive eigenvalues and $q$ negative eigenvalues, then we
have $p+q=n$ and $tr M=p-q$. At the same time, we introduce the
corresponding {\it dual\/} $K$-matrix $K^+(u)$ which is a generic
solution of the dual RE (\ref{DRE-V})
\begin{eqnarray}
K^+(u)=\bar \xi -(u+\frac{n}{2}\eta) \bar M, \quad \bar
M^2=1,\label{km+++}
\end{eqnarray}
where $\bar \xi$ is a boundary parameter and $\bar M$ is an $n\times
n$ boundary parameter dependent matrix, whose eigenvalues are $\pm
1$. Again, we suppose that there are $\bar p$ positive eigenvalues
and $\bar q$ negative eigenvalues, then we have $\bar p + \bar q=n$
and $tr \bar M=\bar p-\bar q$. Besides the dual RE, the $K$-matrix
also satisfies the following properties \bea
 K^+(-\frac{n}{2}\eta)=\bar{\xi},\quad K^+(u)=-u\,\bar{M}+\ldots,\quad u\rightarrow\infty.\label{K+properties}
\eea

The Hamiltonian of the open spin chain specified by the $K$-matrices
$K^{\pm}(u)$ (\ref{km-}) and (\ref{km+++}) can be expressed in terms
of  the transfer matrix (\ref{trans}) as
\begin{eqnarray}
&&H=\eta \frac{\partial \ln t(u)}{\partial
u}|_{u=0,\theta_j=0} \nonumber \\
&&\quad = 2\sum_{j=1}^{N-1}P_{j,j+1} + \eta \frac{tr_0
{K^+_0}^\prime(0)}{tr_0 K^+_0(0)} +2 \frac{tr_0 K_{0}^+(0)
P_{01}}{tr_0 K^+_0(0)} +\eta \frac{1}{\xi}{K_{N}^-}^\prime(0).
\label{oh}
\end{eqnarray}

\subsection{Operator product identities}

Similar to the closed spin chain case in the previous section, we
apply the fusion technique to study the open spin chain. In this
case, we need to use the fusion techniques both  for $R$-matrices
\cite{Kar79} and for $K$-matrices \cite{Mez92,Zho96}. We only
consider the antisymmetric fusion procedure which leads to the
desired operator identities to determine the spectrum of the
transfer matrix $t(u)$ given by (\ref{trans}).

Following \cite{Mez92,Zho96}, let us introduce the fused
$K$-matrices and double-row monodromy matrices by the following
recursive relations
\begin{eqnarray}
K_{ 1, \ldots, m }^+(u)&=&K_{\langle 2, \ldots, m \rangle}^+(u-\eta)R_{1m}(-2u-n\eta+(m-1)\eta) \ldots\no\\
&&\quad \times R_{12}(-2u-n\eta+\eta) K_{1}^+(u), \label{K-matrix-fusion-1} \\
K_{ \langle 1, \ldots, m \rangle}^+(u)&=&P^{(-)}_{1,\ldots, m} K_{ 1, \ldots, m }^+(u)P_{1, \ldots, m}^{(-)}, \\
K^-_{1, \ldots, m }(u)&=&K^-_{1}(u)R_{21}(2u-\eta)\ldots R_{m1}(2u-(m-1)\eta)
K^-_{\langle 2, \ldots, m \rangle}(u-\eta),\\
K_{ \langle 1, \ldots, m \rangle }^-(u)&=&P^{(-)}_{1,\ldots, m}\, K_{ 1, \ldots, m }^-(u)\,P_{1, \ldots, m}^{(-)}, \\
{\cal{J}}_{ 1, \ldots, m }(u)&=&
{\cal{J}}_{1}(u)R_{21}(2u-\eta)\ldots R_{m1}(2u-(m-1)\eta)
{\cal{J}}_{\langle 2, \ldots, m \rangle}(u-\eta), \\
{\cal{J}}_{\langle 1, \ldots, m \rangle}(u)&=&P_{1, \ldots, m}^{(-)}
 {\cal{J}}_{ 1, \ldots, m }(u)P_{1, \ldots, m}^{(-)}
=T_{\langle 1, \ldots, m \rangle}(u) K_{\langle 1,
\ldots, m \rangle}^-(u) \hat T_{\langle 1, \ldots, m \rangle}(u),\label{K-matrix-fusion-2}
\end{eqnarray}
where the fused one-row monodromy matrix $T_{\langle 1, \ldots, m \rangle}(u)$ is given by (\ref{Fused-transfer-period}) and
\bea
 \hat{T}_{\langle 1,\ldots,m\rangle}(u)=P^{(-)}_{1,2,\ldots, m}\,\hat{T}_1(u)\hat{T}_2(u-\eta)\ldots \hat{T}_m(u-(m-1)\eta)\,P^{(-)}_{1,2,\ldots, m}.
\eea
For the open spin chain, the $m$-th  fused transfer matrix
$t_m (u)$ constructed by the antisymmetric fusion procedure is given by
\begin{eqnarray}
&&t_m (u)=tr_{ 1, \ldots, m}\{K_{\langle 1, \ldots,
m\rangle}^+(u){\cal{J}}_{\langle 1, \ldots, m\rangle}(u)\},\quad m=1,\ldots,n,\label{trans-fused}
\end{eqnarray}
which includes the fundamental transfer matrix $t(u)$ given by
(\ref{trans}) as the first one, i.e., $t(u)=t_1(u)$. It follows from
the fusion of the $R$-matrix \cite{Kar79}  and that of the
$K$-matrices \cite{Mez92,Zho96} that the fused transfer matrices
constitute commutative families, namely, \bea
[t_i(u),\,t_j(v)]=0,\quad i,j=1,\ldots,n. \eea Moreover, we remark
that $t_n(u)$ is the so-called quantum determinant and that for generic $u$
and $\{\theta_j\}$ it is proportional to the identity operator, namely,
\begin{eqnarray}
t_n(u) &=&\Delta_q(u)\times {\rm id},\\
\Delta_q(u)&=&\Delta_q\{T(u)\}\Delta_q\{\hat
T(u)\}\Delta_q\{K^+(u)\}\Delta_q\{K^-(u)\} \no \\
&=&\prod_{l=1}^N(u-\theta_l+\eta)(u+\theta_l+\eta)
\prod_{l=1}^N\prod_{k=1}^{n-1}(u-\theta_l-k\eta)(u+\theta_l-k\eta)\no\\
&&\times \prod_{i=1}^{n-1}\prod_{j=1}^{i}(2u-(i+j)\eta)(-2u+(n-2-i-j)\eta)
\no \\
&&\times (-1)^{q+\bar q} \prod_{k=0}^{\bar q-1}(-u+\frac{n-2}{2}\eta -\bar
\xi-k\eta) \prod_{k=0}^{\bar p-1}(-u+\frac{n-2}{2}\eta+\bar
\xi-k\eta)
\no \\
&&\times \prod_{k=0}^{q-1}(u-\xi-k\eta)\prod_{k=0}^{p-1}(u+\xi-k\eta). \label{1109-1}
\end{eqnarray}
The commutativity of the transfer matrices with different spectral
parameters implies that they have  common eigenstates. Let
$|\Psi\rangle$ be a common eigenstate of $\{t_m(u)\}$, which does
not depend upon $u$, with the eigenvalue $\Lambda_m(u)$, i.e., \bea
t_m(u)|\Psi\rangle=\Lambda_m(u)|\Psi\rangle,\quad m=1,\ldots
n.\label{Eigenvalue-open} \eea

Now let us evaluate the product of the fundamental transfer matrix
and the fused ones at some special points \bea &&
t(\pm\theta_j)t_m(\pm\theta_j-\eta)=tr_{1\ldots
m+1}\lt\{{\cal{J}}_1^{t_1}(\pm\theta_j)K_1^+(\pm\theta_j)^{t_1}
\right. \no \\
&&\qquad\qquad \qquad\qquad \qquad \left. \times
     {\cal{J}}_{\langle 2,\ldots,m+1\rangle}(\pm\theta_j-\eta)K^+_{\langle 2,\ldots m+1\rangle}(\pm\theta_j-\eta)\rt\}\no\\
     &\stackrel{(\ref{crosing-unitarity})}{=}& \prod_{k=1}^m\rho^{-1}_2(\pm 2\theta_j-k\eta)\times
     tr_{1\ldots m+1}\lt\{{\cal{J}}_1^{t_1}(\pm\theta_j)K_1^+(\pm\theta_j)^{t_1}\rt.\no\\
     &&\times R^{t_1}_{12}(\mp 2\theta_j+\eta-n\eta)\ldots R^{t_1}_{1\,m+1}(\mp 2\theta_j+m\eta-n\eta)\no\\
     &&\times R^{t_1}_{1\,m+1}(\pm 2\theta_j-m\eta)\ldots
     R^{t_1}_{12}(\pm 2\theta_j-\eta)\no\\
     &&\times \lt.{\cal{J}}_{\langle 2,\ldots,m+1\rangle}(\pm\theta_j-\eta)K^+_{\langle 2,\ldots m+1\rangle}(\pm\theta_j-\eta)\rt\}\no\\
&= & \prod_{k=1}^m\rho^{-1}_2(\pm 2\theta_j-k\eta)\times
     tr_{1\ldots m+1}\lt\{\rt.\no\\
     &&\times R_{1\,m+1}(\mp 2\theta_j+m\eta-n\eta)\ldots R_{12}(\mp 2\theta_j+\eta-n\eta)K^+_1(\pm\theta_j)\no\\
     &&\times \lt.{\cal{J}}_{1,\ldots,m+1}(\pm\theta_j)K^+_{\langle 2,\ldots,m+1\rangle}(\pm\theta_j-\eta)\rt\}\no\\
&= &  \prod_{k=1}^m\rho^{-1}_2(\pm 2\theta_j-k\eta)\times
     tr_{1\ldots m+1}\lt\{K^+_{\langle 2,\ldots,m+1\rangle}(\pm\theta_j-\eta)\rt.\no\\
     &&\times R_{1\,m+1}(\mp 2\theta_j+m\eta-n\eta)\ldots R_{12}(\mp 2\theta_j+\eta-n\eta)K^+_1(\pm\theta_j)\no\\
     &&\times \lt.{\cal{J}}_{1,\ldots,m+1}(\pm\theta_j)\rt\}\no\\
&= & \prod_{k=1}^m\rho^{-1}_2(\pm 2\theta_j-k\eta)\times
     tr_{1\ldots m+1}\lt\{K^+_{1,\ldots,m+1}(\pm\theta_j){\cal{J}}_{1,\ldots,m+1}(\pm\theta_j)\rt\}\no\\
 &\stackrel{(\ref{A.2})}{=}&\prod_{k=1}^m\rho^{-1}_2(\pm 2\theta_j-k\eta)\times
     tr_{1\ldots m+1}\lt\{K^+_{1,\ldots,m+1}(\pm\theta_j)P^{(-)}_{1,\ldots,m+1}{\cal{J}}_{1,\ldots,m+1}(\pm\theta_j)\rt\}\no\\
 &\stackrel{(\ref{A.11})}{=}&\prod_{k=1}^m\rho^{-1}_2(\pm 2\theta_j-k\eta)\times
     tr_{1\ldots m+1}\lt\{K^+_{\langle 1,\ldots,m+1\rangle}(\pm\theta_j){\cal{J}}_{1,\ldots,m+1}(\pm\theta_j)P^{(-)}_{1,\ldots,m+1}\rt\}\no\\
 &\stackrel{(\ref{A.10})}{=}&\prod_{k=1}^m\rho^{-1}_2(\pm 2\theta_j-k\eta)\times
     tr_{1\ldots m+1}\lt\{K^+_{\langle 1,\ldots,m+1\rangle}(\pm\theta_j){\cal{J}}_{\langle
     1,\ldots,m+1\rangle}(\pm\theta_j)\rt\}.
\eea According to the definition (\ref{trans-fused}), we thus have
the following functional relations among the transfer matrices
\begin{eqnarray}
&&t(\pm\theta_j)t_m (\pm\theta_j-\eta)=t_{m+1}(\pm\theta_j)
\prod_{k=1}^{m}\rho_2^{-1}(\pm2\theta_j-k\eta), \label{openfun1} \\
&& \qquad \quad j=1, \ldots, N; \quad m=1, \ldots, n-1. \no
\end{eqnarray}
In terms of the corresponding eigenvalues, the above relations
become \bea &&\Lambda(\pm\theta_j)\Lambda_m
(\pm\theta_j-\eta)=\Lambda_{m+1}(\pm\theta_j)
\prod_{k=1}^{m}\rho_2^{-1}(\pm2\theta_j-k\eta), \label{Eig-fuction-t} \\
&& \qquad \quad j=1, \ldots, N; \quad m=1, \ldots, n-1. \no
\eea
One may check that the fused transfer matrices $t_m(u)$ have some zero points, which allows us
to rewrite the transfer matrices as
\begin{eqnarray}
&&t_m (u)
=\prod_{i=1}^{m-1}\prod_{j=1}^{i}(2u-i\eta-j\eta)(-2u+(2m-2-n)\eta-i\eta-j\eta)
\no \\
&&\qquad \qquad\times
\prod_{l=1}^N\prod_{k=1}^{m-1}(u-\theta_l-k\eta)(u+\theta_l-k\eta)\,
\tau_{m}(u). \label{tau-fused}
\end{eqnarray}
Since the operator $\tau_m(u)$ is proportional to the transfer matrix $t_m(u)$ by c-number coefficient, the corresponding eigenvalue $\bar\Lambda_m(u)$
has the following relation with $\Lambda_m(u)$
\bea
&& \Lambda_m(u)=\prod_{i=1}^{m-1}\prod_{j=1}^{i}(2u-i\eta-j\eta)(-2u+(2m-2-n)\eta-i\eta-j\eta)
\no \\
&&\qquad \qquad\times
\prod_{l=1}^N\prod_{k=1}^{m-1}(u-\theta_l-k\eta)(u+\theta_l-k\eta)\bar\Lambda_m(u).\label{Relation-t-tau}
\eea
It follows from the definitions of the fused transfer matrices (\ref{trans-fused}) that the eigenvalue $\bar\Lambda(u)$ of  the resulting
commutative operator $\tau_m(u)$, as a function of $u$,
is a polynomial of degree $2N+2m$.  The functional relations (\ref{openfun1}) give rise to that the eigenvalue $\bar\Lambda_m(u)$ of $\tau_m(u)$ satisfies
the following relations
\begin{eqnarray}
\bar\Lambda(\pm\theta_j)\bar\Lambda_m (\pm\theta_j-\eta)&=&\bar\Lambda_{m+1}(\pm\theta_j)
\prod_{k=1}^{m}
\rho_2^{-1}(\pm2\theta_j-k\eta)\rho_0(\pm\theta_j),  \label{Eig-funtion-tau}\\
&&m=1,\ldots,n-1,\quad j=1,\ldots,N,\no
\end{eqnarray}
where the function $\rho_0(u)$ is given by
\begin{eqnarray}
 \rho_0 (u)
=\prod_{l=1}^N(u-\theta_l-\eta)(u+\theta_l-\eta)
\prod_{k=2}^m(2u-k\eta)(-2u-k\eta+(n-2)\eta).\no
\end{eqnarray}
Then  $\tau_n(u)$ is  proportional to identity operator with a known
coefficient $\bar\Lambda_n(u)$
\begin{eqnarray}
\hspace{-0.8truecm}\bar\Lambda_n(u) &=&\hspace{-0.28truecm}\prod_{l=1}^N(u-\theta_l+\eta)(u+\theta_l+\eta)
\prod_{k=0}^{\bar q-1}(-u+\frac{n-2}{2}\eta -\bar \xi-k\eta)
\no \\
\hspace{-0.8truecm}&& \hspace{-0.28truecm}\times (-1)^{q+\bar q}\prod_{k=0}^{\bar
p-1}(-u+\frac{n-2}{2}\eta+\bar
\xi-k\eta)\prod_{k=0}^{q-1}(u-\xi-k\eta)
\prod_{k=0}^{p-1}(u+\xi-k\eta). \label{Deterinmant-open}
\end{eqnarray}

\subsection{Asymptotic behaviors of the transfer matrices}
The definitions (\ref{K-matrix-fusion-1})-(\ref{trans-fused}) of the
fused $K$-matrices, the fused monodromy matrices and the fused
transfer matrices and the asymptotic behaviors (\ref{K-properties})
and (\ref{K+properties}) imply that the asymptotic behaviors of the
operators $\{\tau_m(u)\}$ given by (\ref{tau-fused}) is completely
fixed by the eigenvalues of the product matrix  $\bar M M$ (see
(\ref{Asymptotic-open}) below). Firstly let us give some properties
of the eigenvalues of $\bar M M$. Suppose
$\{\lambda_l|l=1,\ldots,n\}$ be the eigenvalues. The fact that
$M^2={\bar M}^2=1$ allows one to derive the following relations
among the eigenvalues,
\begin{eqnarray}
\sum_{l=1}^n\lambda_l^k=tr\{(\bar M M)^k\}=tr \{(M\bar M)^k\} =tr\{(\bar M
M)^{-k}\}=\sum_{l=1}^n\lambda_l^{-k},
\quad \forall k. \label{10-154-2}
\end{eqnarray} Meanwhile we know that
\begin{eqnarray}
{\rm Det}|\bar M M|=\lambda_1\ldots\lambda_n=(-1)^{q+\bar q}.
\end{eqnarray}
This implies that the eigenvalues of $M\bar M$ should take the following form
\bea
 \{\lambda_1, \ldots, \lambda_n\} = \{1, \ldots, 1, -1, \ldots, -1,
e^{-i \vartheta_1}, e^{i \vartheta_1}, \ldots, e^{-i \vartheta_r},
e^{i \vartheta_r}\}, \label{4.3-2}
\eea  where $\vartheta_j$ are some continuous free parameters which are related to boundary
interaction terms (e.g., the boundary magnetic fields).
The maximum number of the continuous  parameters is $n/2$ if $n$ is even and is $(n-1)/2$ if $n$ is odd.

Some remarks are in order. When $M$ and $\bar M$ commute with each
other and thus can be diagonalized simultaneously by some gauge
transformation, the corresponding open spin chain can be
diagonalized by the algebraic Bethe ansatz method after a global
gauge transformation \cite{mar}. In case of the boundary parameters
(which are related to the matrices $M$ and $\bar M$) have some
constraints so that a proper ``local vacuum state" exists, the
generalized algebraic Bethe ansatz method \cite{cao,Yan04} can be
used to obtain the Bethe ansatz solutions of the associated open
spin chains \cite{Yan05,Nep09}. However, the results in \cite{cao1}
strongly suggest that for generic $M$ and $\bar M$ such a simple
``local vacuum state" do not exist even for the $su(2)$ case.

The asymptotic behaviors (\ref{K-properties}) and (\ref{K+properties}) enable us to derive that
the eigenvalue $\bar\Lambda_m(u)$ of  the operators $\{\tau_m(u)\}$ given by (\ref{tau-fused}) have the following asymptotic behaviors
\bea
 \bar\Lambda_m(u)=(-1)^m\delta_m\,u^{2N+2m}+\ldots,\quad m=1,\ldots,n,\quad u\rightarrow\infty,\label{Asymptotic-open}
\eea where
\bea
 \delta_m=\sum_{1\leq i_1 <i_2\ldots<i_m\leq n}\lambda_{i_1}\ldots \lambda_{i_m},\quad m=1,\ldots,n.
\eea Keeping the fact that $\bar\Lambda_n(u)$ has been already fixed
(\ref{Deterinmant-open}) in the mind, we need to determine the
eigenvalues of the other $n-1$ transfer matrices
$\{\tau_m(u)|m=1,\ldots,n-1\}$. It is also known from
(\ref{tau-fused}) that $\bar\Lambda_m(u)$ , as a function of $u$, is
a polynomial of degree $2N+2m$. Thanks to the very functional
relations (\ref{Eig-funtion-tau}) and the asymptotic behaviors
(\ref{Asymptotic-open}), one can completely determine the
eigenvalues of the transfer matrix and the other higher fused
transfer matrices by providing some other values of the eigenvalue
functions at $\sum_{m=1}^{n-1}2m$ special points (e.g. see
(\ref{t1-1})-(\ref{t2-4}) or (\ref{t-4-1})-(\ref{t-4-12}) below).
The method has been proven in \cite{cao1} to be successful in
solving the open spin chains related to  $su(2)$ algebra. In the
following section, we shall apply the method to solve the open spin
chains associated with $su(n)$ algebra.

For this purpose, let us first factorize out the contributions of
$K$-matrices which are relevant to the quantum determinant
$\bar\Lambda_n(u)$ (\ref{Deterinmant-open}) by introducing $n$
functions $\{K^{(l)}(u)|l=1,\ldots,n\}$ which are polynomials of $u$
with a degree 2. The functions depend only on  the boundary
parameters $\xi$ and $\bar\xi$ and satisfy the following relations
\begin{eqnarray}
&& \prod_{l=1}^n K^{(l)}(u-(l-1)\eta) = (-1)^{q+\bar q}
\prod_{k=0}^{\bar
q-1}(-u+\frac{n-2}{2}\eta -\bar \xi-k\eta) \no \\
&& \quad\quad\quad\quad \times \prod_{k=0}^{\bar p-1}(-u+\frac{n-2}{2}\eta+\bar
\xi-k\eta)\prod_{k=0}^{q-1}(u-\xi-k\eta)
\prod_{k=0}^{p-1}(u+\xi-k\eta), \label{keyk2}\\[6pt]
&&K^{(l)}(u) K^{(l)}(-u-l\eta)=K^{(l+1)}(u) K^{(l+1)}(-u-l\eta), \quad
l=1, \cdots, n-1. \label{keyk1}
\end{eqnarray} From the solution to the above equations, one can construct a nested T-Q ansatz for the
eigenvalues $\Lambda_m(u)$. It is remarked that there are some
different solutions to the above equations. However, it was shown in
\cite{cao, Nep13-1} that for the $su(2)$ open spin chain any choice
of the above equation leads to a complete set of solutions of the
the corresponding model. It is believed that different choices of
the solution might only give rise to different parameterizations of
the eigenvalues.
\section{$su(3)$-invariant spin chain with non-diagonal boundary term}
\setcounter{equation}{0}


In this section, we use the method outlined in the previous section
to give the Bethe ansatz solution of the $su(3)$-invariant spin
chain with generic  boundary terms.  Without loss of generality, we
take the corresponding $M$ and $\bar M$ with  $p=\bar p=1$ and the
eigenvalues of $\bar M M$ being \bea
 (\lambda_1,\lambda_2,\lambda_3)=(1,e^{-i \vartheta}, e^{i \vartheta}),
\eea as an example to demonstrate our method in detail.

The functional relations (\ref{Eig-funtion-tau}) of the eigenvalues
$\bar\Lambda_m(u)$ now read
\begin{eqnarray}
\bar\Lambda(\pm\theta_j)\bar\Lambda_m (\pm\theta_j-\eta)=\bar\Lambda_{m+1}(\pm\theta_j)
\prod_{k=1}^{m}
\rho_2^{-1}(\pm2\theta_j-k\eta)\rho_0(\pm\theta_j), \, m=1,2,\, j=1,\ldots,N,\label{Eig-funtion-tau-su(3)}
\end{eqnarray}
where the function $\rho_0(u)$ is given by
\begin{eqnarray}
 \rho_0 (u)
&=&\prod_{l=1}^N(u-\theta_l-\eta)(u+\theta_l-\eta)
\prod_{k=2}^m(2u-k\eta)(-2u-k\eta+\eta),\no\\
\bar\Lambda_3(u)&=&\prod_{l=1}^N(u-\theta_l+\eta)(u+\theta_l+\eta)\no\\
&&\quad\times(\bar\xi+\frac{\eta}{2}-u)(\xi+u) (\bar\xi+\frac{\eta}{2}+u)(\xi-u)(\bar\xi-\frac{\eta}{2}+u)(\xi-u+\eta).
\end{eqnarray}
Let us introduce 3 functions $\{K^{(l)}|l=1,2,3\}$ as follows
\bea
  K^{(1)}(u)&=& (\bar\xi+\frac{1}{2}\eta-u)(\xi+u),\label{K-3-1}\\
  K^{(2)}(u)&=& (\bar\xi+\frac{3}{2}\eta+u)(\xi-u-\eta),\\
  K^{(3)}(u)&=& (\bar\xi+\frac{3}{2}\eta+u)(\xi-u-\eta),\label{K-3-2}
\eea which satisfy the following relations
\bea
&&K^{(1)}(u) K^{(2)}(u-\eta)K^{(3)}(u-2\eta)\no\\
&&\quad\quad=(\bar\xi+\frac{\eta}{2}-u)(\xi+u) (\bar\xi+\frac{\eta}{2}+u)(\xi-u)
(\bar\xi-\frac{\eta}{2}+u)(\xi-u+\eta), \label{1217}\\[6pt]
&&K^{(l)}(u) K^{(l)}(-u-l\eta)=K^{(l+1)}(u) K^{(l+1)}(-u-l\eta),
\quad l=1, 2. \label{keyk1-su(3)} \eea From the definitions
(\ref{trans-fused}) of the fused transfer matrices $t_m(u)$ and the
asymptotic behaviors of the $K$-matrices $K^{\pm}(u)$, we have that
the eigenvalues of the transfer matrices have the following
asymptotic behaviors
\begin{eqnarray}
\bar\Lambda(u)|_{u\rightarrow \infty}&=& - tr(\bar M M)u^{2N+2} +\ldots  =-\sum_{i=1}^3 \lambda_{i} u^{2N+2} +\ldots \no \\
&=&-(1 + 2\cos \vartheta)u^{2N+2}+\ldots.
,\label{A-3-1}\\[6pt]
\bar\Lambda_2(u)|_{u\rightarrow \infty} &=& tr_{12}\left\{ P_{1,2}^{(-)}(\bar M M)_1 (\bar
M M)_2P_{1,2}^{(-)}\right\}u^{2N+4}  +\ldots \no \\
&=&  \sum_{1 \leq i_1<i_{2} \leq 3}
\lambda_{i_1}\lambda_{i_2} u^{2N+4} +\ldots \no \\
&=& (2\cos\vartheta +1) u^{2N+4}+\ldots.\label{A-3-2}
\end{eqnarray}
Moreover, the properties of $R$-matrix
(\ref{Initial})-(\ref{Fusion}) and $K$-matrices (\ref{K-properties})
and (\ref{K+properties}) allow us to derive that the fused transfer
matrices satisfy the following properties at some special points:
\begin{eqnarray}
&&t(0)= (-1)^N \xi \prod_{l=1}^N
(\theta_l+\eta)(\theta_l-\eta)tr \{K^+(0)\}\,\times{\rm id},\label{t1-1} \\
&&t(-\frac32\eta)=(-1)^N \bar \xi \prod_{l=1}^N
(\theta_l+\frac32\eta)(\theta_l-\frac32\eta)tr
\{K^-(-\frac32\eta)\}\,\times {\rm id}, \label{t1-2}\\
&& t_2(\frac{\eta}{2})=tr_{12}\left\{ P_{12}^-
K_{2}^+(-\frac{\eta}{2})R_{12}(-3\eta)K_{1}^+(\frac{\eta}{2})P_{12}^-\right\}\left(\frac{\eta^2}{4}-\xi^2\right)\eta
\no \\
&&\qquad\qquad \quad \times \prod_{l=1}^N
(\theta_l+\frac32\eta)(\theta_l-\frac32\eta)
(\theta_l+\frac{\eta}{2})(\theta_l-\frac{\eta}{2})\,\times {\rm id}, \label{t2-2} \\
&& t_2(-\eta)=tr_{12}\left\{
P_{12}^-K_{1}^-(-\eta)R_{21}(-3\eta)K_{2}^-(-2\eta)P_{12}^-\right\}\left(\frac{\eta^2}{4}-\bar
\xi^2\right)\eta
\no \\
&&\qquad\qquad \quad \times \prod_{l=1}^N
(\theta_l+\eta)(\theta_l-\eta) (\theta_l+2\eta)(\theta_l-2\eta)\,\times {\rm id},
\label{t2-3}\\
&& t_2(0)= (-1)^N 2 \xi \eta^2 \prod_{l=1}^N
(\theta_l+\eta)(\theta_l-\eta)tr \{K^+(0)\}\,  t(-\eta),\label{t2-1} \\
&& t_2(-\frac{\eta}{2})= (-1)^N 2 \bar \xi \eta^2 \prod_{l=1}^N
(\theta_l+\frac32\eta)(\theta_l-\frac32\eta)tr \{K^-(-\frac32\eta)\}
\,t(-\frac{\eta}{2}),\label{t2-4}
\end{eqnarray} The above relations allow us to derive similar relations of the eigenvalues $\{\bar\Lambda_m(u)\}$. Then
the resulting relations (total number of the conditions is equal to
$2+4=6$), the very relations (\ref{Eig-funtion-tau-su(3)}) and the
asymptotic behaviors (\ref{A-3-1})-(\ref{A-3-2}) allow us to
determine the eigenvalues $\bar\Lambda_m(u)$ (also $\Lambda_m(u)$
via the relations (\ref{Relation-t-tau})).

Let us define the corresponding  $Q^{(r)}(u)$  for the open spin chains
\begin{eqnarray}
&& Q^{(0)}(u)=\prod_{j=1}^N(u-\theta_j)(u+\theta_j), \label{10-14-3-1118} \\
&&Q^{(r)}(u)=\prod_{l=1}^{L_r}(u-\lambda^{(r)}_l)(u+\lambda^{(r)}_l+r\eta),
\quad r=1, \ldots, n-1,\label{Q-functions-Open-1}\\
&&Q^{(n)}(u)=1,\label{Q-functions-Open-2}
\end{eqnarray} where $\{L_r|r=1,\ldots n-1\}$ are some non-negative integers. In the following part of the paper, we adopt the convention
\bea a(u)=Q^{(0)}(u+\eta),\quad d(u)=Q^{(0)}(u). \eea In order to
construct the solution of open $su(3)$ spin chain, we introduce
three $\tilde z(u)$ functions
\begin{eqnarray}
&&\tilde z_{1}(u)=z_{1}(u)+x_{1}(u), \quad \tilde z_{2}(u)=z_{2}(u),
\quad \tilde z_{3}(u)=z_{3}(u).
\end{eqnarray}
Here $z_{m}(u)$ is defined as
\begin{eqnarray}
z_{m}(u) =\frac{u(u+\frac 32 \eta)}{(u+\frac{(m-1)}{2}\eta)(u+\frac
{m}{2}\eta)}K^{(m)}(u)d(u)
\frac{Q^{(m-1)}(u+\eta)Q^{(m)}(u-\eta)}{Q^{(m-1)}(u)Q^{(m)}(u)},\quad
m=1,2,3, \label{2z2-1118-1216}
\end{eqnarray}
and $x_1(u)$ is defined as
\begin{eqnarray}
x_{1}(u)= u(u+ \frac 3 2\eta)a(u)d(u) \frac{F_{1}(u)}{Q^{(1)}(u)}.
\end{eqnarray}
The nested functional T-Q ansatz  is expressed as
\begin{eqnarray}
&& \Lambda(u)=\sum_{i_1=1}^{3} \tilde z_{i_1}(u) =\sum_{i_1=1}^{3}
z_{i_1}(u) +u(u+ \frac 32\eta)a(u)d(u) \frac{F_{1}(u)}{Q^{(1)}(u)}.
\label{4t1-103-1107-1216}
\end{eqnarray}
\begin{eqnarray}
\Lambda_2(u)=\sum_{1\leq i_1<i_2 \leq 3 } \tilde z_{i_1}(u) \tilde
z_{i_2}(u-\eta) - x_1(u) z_2 (u-\eta).\label{t2-3-2}
\end{eqnarray}
Here $F_1(u)$ is a polynomial of degree $2L_1-2N$. The consistency
of zero residues of $\Lambda(u)$ at $\lambda_j^{(1)}$ and
$-\lambda_j^{(1)}-\eta$ requires
\begin{eqnarray}
F_1(u) =f_1(u)Q^{(2)}(-u-\eta),
\end{eqnarray}
with
\begin{eqnarray}
f_1(u) = f_1(-u-\eta). \label{f11-103-1-1106-1216}
\end{eqnarray}
Let all terms with $f_1(u)$ in $\Lambda_m(u)$ be zero at all the
degenerate points considered in (\ref{t1-1}-\ref{t2-4}). $f_1(u)$
can be given by
\begin{eqnarray}
&&f_1(u)=  c u(u+\frac{1}{2}\eta)^2 (u+\eta),
\end{eqnarray}
where $c$ is a constant. This allows us to write down the explicit
nested T-Q ansatz (\ref{4t1-103-1107-1216})-(\ref{t2-3-2}) as follows
\bea
 \Lambda(u)&=&\frac{2u+3\eta}{2u+\eta} K^{(1)}(u)a(u)
\frac{Q^{(1)}(u-\eta)}{Q^{(1)}(u)}
\nonumber \\
&&\,+
\frac{2u(2u+3\eta)}{(2u+\eta)(2u+2\eta)}K^{(2)}(u)d(u)\frac{Q^{(1)}(u+\eta)Q^{(2)}(u-\eta)}{Q^{(1)}(u)Q^{(2)}(u)}
\nonumber \\
&&\, + \frac{2u}{2u+2\eta} K^{(3)}(u)d(u)
\frac{Q^{(2)}(u+\eta)}{Q^{(2)}(u)}\no\\
&&\, +c\,u(u+\frac 32\eta)a(u)d(u)
\frac{u(u+\frac{1}{2}\eta)^2(u+\eta)Q^{(2)}(u-\eta)}{Q^{(1)}(u)},
\label{3t1-101}\eea \bea
\Lambda_2(u)&=&\rho_2(2u-\eta)d(u-\eta)\left\{
\frac{(2u-2\eta)(2u+3\eta)}{2u(2u-\eta)}
K^{(1)}(u)K^{(2)}(u-\eta)a(u)
\frac{Q^{(2)}(u-2\eta)}{Q^{(2)}(u-\eta)}
\right.\nonumber \\
&& \, \left. +
\frac{(2u-2\eta)(2u+3\eta)}{(2u+\eta)2u}K^{(1)}(u)K^{(3)}(u-\eta)a(u)\frac{Q^{(1)}(u-\eta)Q^{(2)}(u)}{Q^{(1)}(u)Q^{(2)}(u-\eta)}
\right.\nonumber \\
&&\, + \frac{(2u-2\eta)(2u+3\eta)}{(2u+\eta)(2u+2\eta)}
K^{(2)}(u)K^{(3)}(u-\eta)d(u)
\frac{Q^{(1)}(u+\eta)}{Q^{(1)}(u)}\no \\
&& \,
+\left.c\,(u-\eta)(u+\frac{3}{2}\eta)a(u)d(u)\frac{u(u+\frac{1}{2}\eta)^2(u+\eta)Q^{(2)}(u)K^{(3)}(u-\eta)}{
Q^{(1)}(u)}\right\}, \label{3t2-101} \eea where the non-negative
integers $L_1$ and $L_2$ satisfy the relation: \bea
L_1=N+L_2+2,\label{LN-1} \eea the functions $\{K^{(l)}(u)|l=1,2,3\}$
are given by (\ref{K-3-1})-(\ref{K-3-2}) and the parameter $c$ is
given by
\begin{eqnarray}
c=2(\cos \vartheta-1).\label{r2}
\end{eqnarray} The above relation and the relations (\ref{Relation-t-tau}) between $\bar\Lambda_m(u)$ and $\Lambda_m(u)$
lead to  that the asymptotic behaviors
(\ref{A-3-1})-(\ref{A-3-2}) of the eigenvalues $\bar\Lambda_m(u)$ are automatically satisfied. Noticing that
\bea
 a(\theta_j-\eta)=d(\theta_j)=0,
\eea one can easily show that the ansatz
(\ref{3t1-101})-(\ref{3t2-101}) also make  the very functional
relations (\ref{Eig-funtion-tau-su(3)}) fulfilled. The regular
property of $\Lambda(u)$  leads to the associated Bethe ansatz
equations, \bea
&&1+\frac{\lambda_l^{(1)}}{\lambda_l^{(1)}+\eta}\frac{K^{(2)}({\lambda_l^{(1)}})d(\lambda^{(1)}_l)}{K^{(1)}({\lambda_l^{(1)}})a(\lambda^{(1)}_l)}
\frac{Q^{(1)}(\lambda^{(1)}_l+\eta)Q^{(2)}(\lambda^{(1)}_l-\eta)}{Q^{(1)}(\lambda^{(1)}_l-\eta)Q^{(2)}(\lambda^{(1)}_l)}\no\\
&&\quad\quad=-c\,\frac{(\lambda^{(1)}_l)^2(\lambda^{(1)}_l+\frac{1}{2}\eta)^3(\lambda^{(1)}+\eta)d(\lambda^{(1)}_l)Q^{(2)}(\lambda^{(1)}_l-\eta)}
{K^{(1)}(\lambda^{(1)}_l)Q^{(1)}(\lambda^{(1)}_l-\eta)},\quad l=1,\ldots, L_1,\label{BAE-Open-3-1}\\[6pt]
&&\frac{\lambda_l^{(2)}+\frac{3}{2}\eta}{\lambda_l^{(2)}+\frac{1}{2}\eta}\frac{K^{(2)}({\lambda_l^{(2)}})}{K^{(3)}({\lambda_l^{(2)}})}
\frac{Q^{(1)}(\lambda^{(2)}_l+\eta)Q^{(2)}(\lambda^{(2)}_l-\eta)}{Q^{(1)}(\lambda^{(2)}_l)Q^{(2)}(\lambda^{(2)}_l+\eta)}=-1,
\quad l=1,\dots L_2.\label{BAE-Open-3-2} \eea One may check that the
chosen $F_1(u)$ and the BAEs
(\ref{BAE-Open-3-1})-(\ref{BAE-Open-3-2}) also guarantee the
regularity of the ansatz $\Lambda_2(u)$  given by (\ref{3t2-101}).
Moreover, the ansatz (\ref{3t1-101})-(\ref{3t2-101}) indeed satisfy
the relations (\ref{t1-1})-(\ref{t2-4}). Finally, we conclude that
the ansatz $\Lambda_m(u)$ given by (\ref{3t1-101})-(\ref{3t2-101})
are the eigenvalues of the transfer matrices $t_m(u)$ of the
$su(3)$-invariant open spin chain with the most general non-diagonal
boundary terms.

The eigenvalue of the Hamiltonian (\ref{oh}) in the case of $n=3$ is given by
\begin{eqnarray}
E= \sum_{l=1}^{L_1}
\frac{2\eta^2}{\lambda_l^{(1)}(\lambda_l^{(1)}+\eta)}
+2(N-1)+\eta\frac{\bar \xi +\frac 3 2 \eta -\bar p\eta
-\xi}{\xi(\bar \xi +\frac 3 2 \eta -\bar p\eta)}+\frac 2 3,
\end{eqnarray}
where the parameters $\{\lambda^{(1)}_l\}$ are the roots of the BAEs
(\ref{BAE-Open-3-1})-(\ref{BAE-Open-3-2}) in the homogeneous limit
$\theta_j=0$.

\section{$su(4)$-invariant spin chain with non-diagonal boundary magnetic fields}
\setcounter{equation}{0}

In this section, we use the method outlined in  Section 3 to give
the Bethe ansatz solution of the  $su(4)$-invariant open  spin chain
with generic boundary terms. The model may include two free
continuous parameters $\vartheta_1$ and $\vartheta_2$ defined in
(\ref{4.3-2}), which is the first non-trivial case to study the
multi-components models beyond $su(2)$ case. Without loss of
generality, we consider the matrices $M$ and $\bar M$ with  $p=2$
and $\bar p=2$.

Let us introduce 4 functions $\{K^{(l)}(u)|l=1,\ldots 4\}$
\begin{eqnarray}
&&K^{(1)}(u)=(\xi+u)(\bar \xi -u),\label{K-4-1} \\
&&K^{(2)}(u)=(\xi+u)(\bar \xi -u), \\
&&K^{(3)}(u)=(\xi-u-2\eta)(\bar \xi +u+2\eta), \\
&&K^{(4)}(u)=(\xi-u-2\eta)(\bar \xi +u+2\eta),\label{K-4-2}
\end{eqnarray} which satisfy the relations (\ref{keyk2})-(\ref{keyk1}) with $n=4$.
Let us  consider the most
general  case in which the eigenvalues of the $\bar M M$ are
\bea
(\lambda_1, \lambda_2,\lambda_3,\lambda_4)=(e^{i\vartheta_1},e^{-i\vartheta_1},e^{i\vartheta_2},e^{-i\vartheta_2}).\label{Eigen-M-4}
\eea
Then the asymptotic behaviors of the eigenvalues of the transfer matrices
read
\begin{eqnarray}
\bar \Lambda_1(u)|_{u\rightarrow \infty} &=& - tr\{\bar M M\}u^{2N+2} +\ldots
=-(2\cos \vartheta_1+2\cos\vartheta_2)u^{2N+2}+\ldots,\label{asym-4-1}\\[6pt]
\bar\Lambda_2(u)|_{u\rightarrow \infty} &=& tr_{12}\left\{
P_{1,2}^{(-)}(\bar M M)_1 (\bar
M M)_2P_{1,2}^{(-)}\right\} u^{2N+4} +\ldots \no \\
&=& \sum_{1 \leq i_1<i_{2} \leq 3}
\lambda_{i_1}\lambda_{i_2}  u^{2N+4} +\ldots \no \\
&=&(2 +4\cos\vartheta_1\cos\vartheta_2) u^{2N+4}+\ldots,\label{asym-4-2}\\[6pt]
\bar\Lambda_3(u)|_{u\rightarrow \infty} &=&-tr_{123}\left\{
P_{1,2,3}^{(-)}(\bar M M)_1 (\bar
M M)_2 (\bar
M M)_2 P_{1,2,3}^{(-)}\right\} u^{2N+6} +\ldots \no \\
&=& -\sum_{1 \leq i_1<i_{2}<i_3 \leq 4}
\lambda_{i_1}\lambda_{i_2}\lambda_{i_3}  u^{2N+6} +\ldots \no \\
&=&-(2\cos\vartheta_1+2\cos\vartheta_2)
u^{2N+6}+\ldots.\label{asym-4-3}
\end{eqnarray}
Moreover, we can derive the following relations among the fused transfer matrices at
some special points:
\begin{eqnarray}
t(0)&=& (-1)^N \xi \prod_{l=1}^N
(\theta_l+\eta)(\theta_l-\eta)tr\{K^+(0)\}\times {\rm id},\label{t-4-1} \\
t(-2\eta)&=& (-1)^N \bar \xi \prod_{l=1}^N
(\theta_l+\frac32\eta)(\theta_l-2\eta)tr\{K^-(-2\eta)\}\times {\rm id}, \label{t-4-2}\\
t_2(0)&=& 3 (-1)^N \xi\eta^2 \prod_{l=1}^N
(\theta_l+\eta)(\theta_l-\eta)tr\{K^+(0)\} \,t_1(-\eta), \label{t-4-3}\\
t_2(\frac{\eta}{2})&=& tr_{12}\lt\{K_{\langle 12 \rangle}^+(\frac{\eta}{2})\rt\}\eta (\frac{\eta^2}{4}-\xi^2)\no\\
&&\quad\times\prod_{l=1}^N (\theta_l-\frac{\eta}
{2})(\theta_l+\frac{\eta}{2})
(\theta_l-\frac32\eta)(\theta_l+\frac32\eta)\,\times {\rm id}, \label{t-4-4}\\
t_2(-\frac{3}{2}\eta)&=& \eta (\frac{\eta^2}{4}-{\bar \xi}^2)
\prod_{l=1}^N (\theta_l-\frac 52\eta)(\theta_l+\frac 52\eta)
(\theta_l-\frac32\eta)(\theta_l+\frac32\eta)\no\\
&&\quad\times tr_{12}\{K_{\langle 12 \rangle }^-(-\frac 32 \eta)\}\,\times{\rm id},\label{t-4-5} \\
t_2(-\eta)&=& 3 (-1)^N \bar \xi\eta^2 \prod_{l=1}^N
(\theta_l+2\eta)(\theta_l-2\eta)tr \{K^-(2\eta)\}\,
t_1(-\eta),\label{t-4-6}
\end{eqnarray}
and
\begin{eqnarray}
t_3(0)&=&12 (-1)^N \xi\eta^4 \prod_{l=1}^N
(\theta_l+\eta)(\theta_l-\eta)tr \{K^+(0)\} \, t_2(-\eta),\label{t-4-7}  \\
t_3(0)&=&12 (-1)^N \bar \xi\eta^4 \prod_{l=1}^N
(\theta_l+2\eta)(\theta_l-2\eta)tr \{K^-(-2\eta)\}\,  t_2(0), \label{t-4-8} \\
t_3(\frac{\eta}{2})&=&12  tr_{12}\{K_{\langle 12 \rangle }^+(\frac{\eta}
{2})\}\eta^5 (\frac{\eta^2}{4}-\xi^2) \, t_1(-\frac32\eta)\no  \\
&& \quad \times \prod_{l=1}^N (\theta_l-\frac{\eta}
{2})(\theta_l+\frac{\eta}{2})
(\theta_l-\frac32\eta)(\theta_l+\frac32\eta), \label{t-4-9} \\
t_3(-\frac{\eta}{2})&=& 12 \eta^5 (\frac{\eta^2}{4}-{\bar \xi}^2)
tr_{23}\{K_{\langle 23 \rangle }^-(-\frac 32 \eta)\}\,t_1(-\frac{\eta}{2})\no \\
&&\quad \times \prod_{l=1}^N (\theta_l-\frac
52\eta)(\theta_l+\frac 52\eta)
(\theta_l-\frac32\eta)(\theta_l+\frac32\eta),\label{t-4-10}
\end{eqnarray}
\begin{eqnarray}
\frac{\partial}{\partial u}  t_3(u)\lt|_{u=\eta}\rt.&=&4\xi \eta^2 (\xi^2-
\eta^2) (-1)^N tr_{123}\{K_{\langle 123 \rangle }^+( \eta)\}\no \\
&&  \times \prod_{l=1}^N \theta^2_l(\theta_l-\eta)(\theta_l+\eta)
(\theta_l-2\eta)(\theta_l+2\eta)\,\times{\rm id},\label{t-4-11}\\
\frac{\partial}{\partial u}  t_3(u)\lt|_{u=-\eta}\rt.&=&4\bar \xi \eta^2 (
\eta^2-{\bar \xi}^2) (-1)^N tr_{123}\{K_{\langle 123 \rangle }^-(- \eta)\}\no \\
&& \times \prod_{l=1}^N (\theta_l-\eta)(\theta_l+\eta)
(\theta_l-2\eta)(\theta_l+2\eta)(\theta_l-3\eta)(\theta_l+3\eta)\,\times{\rm id}.\label{t-4-12}
\end{eqnarray}
The above relations allow us to derive similar relations of the
eigenvalues $\{\bar\Lambda_m(u)|m=1,2,3\}$. Then the resulting
relations (total number of the conditions is equal to $2+4+6=12$),
the very relations (\ref{Eig-funtion-tau}) with $n=4$ and the
asymptotic behaviors (\ref{asym-4-1})-(\ref{asym-4-3}) allow us to
determine the eigenvalues $\bar\Lambda_m(u)$ (also $\Lambda_m(u)$
via the relations (\ref{Relation-t-tau})).

For the $su(4)$ open  spin chain, the $\tilde z(u)$ functions are
\begin{eqnarray}
&&\tilde z_{2l-1}(u)=z_{2l-1}(u)+x_{2l-1}(u),\quad l=1, 2, \\
&&\tilde z_{2l}(u)=z_{2l}(u), \quad l=1, 2.
\end{eqnarray}
Here the function $z_{m}(u)$ is defined in (\ref{2z2-1118-1216})
with $m=1, \cdots, 4$ and $Q^{(4)}(u)\equiv 1$. The function
$x_m(u)$ is
\begin{eqnarray}
x_{2l-1}(u)= u(u+ 2\eta)a(u)d(u) \frac{F_{2l-1}(u)}{Q^{(2l-1)}(u)},
\quad l=1, 2.
\end{eqnarray}
The nested T-Q ansatz can be constructed as
\begin{eqnarray}
&&\Lambda(u)=\sum_{i_1=1}^{4} \tilde z_{i_1}(u) =\sum_{i_1=1}^{4}
z_{i_1}(u) +u(u+ 2\eta)a(u)d(u) \sum_{l=1}^2
\frac{F_{2l-1}(u)}{Q^{(2l-1)}(u)}. \label{4t1-103-1107}
\end{eqnarray}
\begin{eqnarray}
&& \Lambda_2(u)=\sum_{1\leq i_1<i_2 \leq 4 } \tilde z_{i_1}(u)
\tilde z_{i_2}(u-\eta) - x_1(u) z_2 (u-\eta) - x_3(u) z_4(u-\eta),
\\
&&\Lambda_3(u)=\sum_{1\leq i_1<i_2 \leq 4 } \tilde z_{i_1}(u) \tilde
z_{i_2}(u-\eta) - x_1(u) z_2 (u-\eta) (\tilde z_3+\tilde z_4)
\nonumber \\
&&\qquad \qquad  -(\tilde z_1+\tilde z_2) x_3(u)z_4(u-\eta).
\end{eqnarray}
With the similar analysis used for the $su(3)$ case, we have
\begin{eqnarray}
&&F_1(u) =f_1(u) K^{(1)}(u)Q^{(2)}(-u-\eta), \label{f111-103-2} \\
&&F_3(u) = f_3(u)K^{(3)}(u)a(-u-3\eta)Q^{(2)}(-u-3\eta),
\label{f111-103-3}
\end{eqnarray}
with
\begin{eqnarray}
&& f_1(u) = f_1(-u-\eta), \label{f11-103-2-1106} \\
&& f_3(u) = f_3(-u-3\eta). \label{f11-103-2-4t1}
\end{eqnarray}
The eigenvalues of the transfer matrices of the $su(4)$-invariant
open chain with the most general non-diagonal boundary terms are
thus given by
\begin{eqnarray}
\Lambda(u)&=&\frac{u+2\eta}{u+\frac{\eta}{2}} K^{(1)}(u)a(u)
\frac{Q^{(1)}(u-\eta)}{Q^{(1)}(u)}
\nonumber \\
&& +
\frac{u(u+2\eta)}{(u+\frac{\eta}{2})(u+\eta)}K^{(2)}(u)d(u)\frac{Q^{(1)}(u+\eta)Q^{(2)}(u-\eta)}{Q^{(1)}(u)Q^{(2)}(u)}
\nonumber \\
&& +
\frac{u(u+2\eta)}{(u+\eta)(u+\frac{3\eta}{2})}K^{(3)}(u)d(u)\frac{Q^{(2)}(u+\eta)Q^{(3)}(u-\eta)}{Q^{(2)}(u)Q^{(3)}(u)}
\nonumber \\
&&+ \frac{u}{u+\frac{3\eta}{2}} K^{(4)}(u)d(u)
\frac{Q^{(3)}(u+\eta)}{Q^{(3)}(u)}
\nonumber \\
&& + u(u+ 2\eta)a(u)d(u)
\left[\frac{F_1(u)}{Q^{(1)}(u)}+\frac{F_3(u)}{Q^{(3)}(u)}\right],
\label{4t1-103}
\end{eqnarray}
\begin{eqnarray}
\Lambda_2(u)&=&\rho_2(2u-\eta)d(u-\eta)\lt\{\frac{(u+2\eta)(u-\eta)(u+\eta)}{(u+\frac{\eta}{2})(u-\frac{\eta}{2})u} K^{(1)}(u)a(u)
K^{(2)}(u-\eta)\rt.\no\\
&&\quad\quad\times \frac{Q^{(2)}(u-2\eta)}{Q^{(2)}(u-\eta)}
\nonumber \\
&&+ \frac{(u+2\eta)(u-\eta)(u+\eta)}{(u+\frac{\eta}{2})u(u+\frac{\eta}{2})}
K^{(1)}(u)a(u)\frac{Q^{(1)}(u-\eta)}{Q^{(1)}(u)}
K^{(3)}(u-\eta)\no\\
&&\quad\quad \times\frac{Q^{(2)}(u)Q^{(3)}(u-2\eta)}{Q^{(2)}(u-\eta)Q^{(3)}(u-\eta)}
\nonumber \\
&&+ \frac{(u+2\eta)(u-\eta)}{(u+\frac{\eta}{2})(u+\frac{\eta}{2})}
K^{(1)}(u)a(u)\frac{Q^{(1)}(u-\eta)}{Q^{(1)}(u)}
K^{(4)}(u-\eta)\frac{Q^{(3)}(u)}{Q^{(3)}(u-\eta)}
\nonumber \\
&& + \frac{(u+2\eta)(u-\eta)}{(u+\frac{\eta}{2})(u+\frac{\eta}{2})}
K^{(2)}(u)d(u)\frac{Q^{(1)}(u+\eta)}{Q^{(1)}(u)}
K^{(3)}(u-\eta) \frac{Q^{(3)}(u-2\eta)}{Q^{(3)}(u-\eta)}
\nonumber \\
&&+\frac{u(u+2\eta)(u-\eta)}{(u+\frac{\eta}{2})(u+\frac{\eta}{2})(u+\eta)}
K^{(2)}(u)d(u)\frac{Q^{(1)}(u+\eta)Q^{(2)}(u-\eta)}
{Q^{(1)}(u)Q^{(2)}(u)}\no\\
&&\quad\quad\times K^{(4)}(u-\eta)
\frac{Q^{(3)}(u)}{Q^{(3)}(u-\eta)}
\nonumber \\
&& + \frac{u(u+2\eta)(u-\eta)}{(u+\frac{3\eta}{2})(u+\frac{\eta}{2})(u+\eta)}
K^{(3)}(u)d(u)\frac{Q^{(2)}(u+\eta)}{Q^{(2)}(u)}
K^{(4)}(u-\eta)
\nonumber \\
&& +\frac{(u+2\eta)}{(u+\frac{\eta}{2})}(u-\eta)(u+\eta)
K^{(1)}(u)a(u)\frac{Q^{(1)}(u-\eta)}{Q^{(1)}(u)} a(u-\eta)
\frac{F_{3}(u-\eta)}{Q^{(3)}(u-\eta)}
\nonumber \\
&& + \frac{(u-\eta)(u+\eta)}{u(u+\frac{\eta}{2})}u(u+2\eta)
 a(u)d(u)\frac{ F_{1}(u)}{Q^{(1)}(u)}
K^{(3)}(u-\eta)
\frac{Q^{(2)}(u)Q^{(3)}(u-2\eta)}{Q^{(2)}(u-\eta)Q^{(3)}(u-\eta)}
\nonumber \\
&&+ u(u+2\eta)(u-\eta)(u+\eta)a(u)d(u)\frac{ F_{1}(u)}{Q^{(1)}(u)}
a(u-\eta) \frac{ F_{3}(u-\eta)}{Q^{(3)}(u-\eta)}
\nonumber \\
&& + \frac{(u-\eta)}{(u+\frac{\eta}{2})}u(u+2\eta)
a(u)d(u)\frac{ F_{1}(u)}{Q^{(1)}(u)}
K^{(4)}(u-\eta) \frac{Q^{(3)}(u)}{Q^{(3)}(u-\eta)}
\nonumber \\
&& + \frac{(u-\eta)(u+\eta)}{(u-\frac{\eta}{2})u}(u-\eta)(u+\eta)
K^{(2)}(u)d(u)\frac{Q^{(1)}(u+\eta)Q^{(2)}(u-\eta)}{Q^{(1)}(u)Q^{(2)}(u)}\no\\
&&\quad\quad\times a(u-\eta)\lt. \frac{
F_{3}(u-\eta)}{Q^{(3)}(u-\eta)}\rt\}, \label{4t2-103}
\end{eqnarray}
\begin{eqnarray}
\Lambda_3(u)&=&\rho_2(2u-\eta)\rho_2(2u-2\eta)\rho_2(2u-3\eta)d(u-\eta)d(u-2\eta)\times\lt\{\rt.\no\\
&&
\frac{(u+2\eta)(u+\eta)(u-2\eta)}{(u+\frac{\eta}{2})(u-\frac{\eta}{2})(u-\frac{\eta}{2})}K^{(1)}(u)a(u)
K^{(2)}(u-\eta)  K^{(3)}(u-2\eta)
\frac{Q^{(3)}(u-3\eta)}{Q^{(3)}(u-2\eta)}
\nonumber \\
&&+ \frac{(u+2\eta)(u-\eta)(u+\eta)(u-2\eta)}{(u+\frac{\eta}{2})u(u-\frac{\eta}{2})(u-\frac{\eta}{2})}
K^{(1)}(u)a(u) K^{(2)}(u-\eta)
K^{(4)}(u-2\eta)\no\\
&&\quad\quad\times \frac{Q^{(2)}(u-2\eta)Q^{(3)}(u-\eta)}{Q^{(2)}(u-\eta)Q^{(3)}(u-2\eta)}
\nonumber \\
&&+ \frac{(u+2\eta)(u-\eta)(u+\eta)(u-2\eta)}{(u+\frac{\eta}{2})u(u+\frac{\eta}{2})(u-\frac{\eta}{2})}
K^{(1)}(u)a(u)  K^{(3)}(u-\eta)
K^{(4)}(u-2\eta)\no\\
&&\quad\quad\times
\frac{Q^{(1)}(u-\eta)Q^{(2)}(u)}{Q^{(1)}(u)Q^{(2)}(u-\eta)}
\nonumber \\
&&+\frac{(u+2\eta)(u-\eta)(u-2\eta)}{(u+\frac{\eta}{2})(u+\frac{\eta}{2})(u-\frac{\eta}{2})}
K^{(2)}(u)d(u)  K^{(3)}(u-\eta) K^{(4)}(u-2\eta)
\frac{Q^{(1)}(u+\eta)}{Q^{(1)}(u)}
\nonumber \\
&&+(u-2\eta)(u+2\eta) \frac{(u+\eta)(u-\eta)}{(u+\frac{\eta}{2})(u-\frac{\eta}{2})}
 K^{(1)}(u)a(u)  K^{(2)}(u-\eta)
\frac{Q^{(2)}(u-2\eta)}{Q^{(2)}(u-\eta)}\no\\
&&\quad\quad\times a(u-2\eta)
\frac{ F_3(u-2\eta)}{Q^{(3)}(u-2\eta)}
\nonumber \\
&&+ (u+2\eta)(u-2\eta)\frac{(u+\eta)(u-\eta)}{(u+\frac{\eta}{2})(u-\frac{\eta}{2})}
a(u)d(u)\frac{ F_1(u)}{Q^{(1)}(u)}\no\\
&&\quad\quad\lt.\times K^{(3)}(u-\eta)\frac{Q^{(2)}(u)}{Q^{(2)}(u-\eta)}
K^{(4)}(u-2\eta)\rt\}, \label{4t3-103}
\end{eqnarray} where
\bea
&&F_1(u)=c_1(\bar\xi-u)(\xi+u)(u-\frac{\eta}{2})u(u+\frac{\eta}{2})^2(u+\eta)(u+\frac{3}{2}\eta)Q^{(2)}(u-\eta),\\
&&F_3(u)=c_3(\bar\xi+u+2\eta)(\xi-u-2\eta)(u+\frac{\eta}{2})(u+\eta)(u+\frac{3}{2}\eta)^2(u+2\eta)(u+\frac{5}{2}\eta)\no\\
&&\quad\quad\times d(u+2\eta)Q^{(2)}(u+\eta). \eea In the above
equation the functions $\{K^{(l)}(u)|l=1,\ldots,4\}$ are given by
(\ref{K-4-1})-(\ref{K-4-2}), the non-negative $\{L_1,L_2,L_3\}$
satisfy the following relation \bea L_1=4+L_2+N,\quad
L_3=4+2N+L_2,\label{L-Constraint-4} \eea and  the parameters $c_1$
and $c_3$ are determined by the eigenvalues (\ref{Eigen-M-4}) of the
corresponding matrix $\bar M M$ through the following equations \bea
\lt\{\begin{array}{l}-4 + c_1+ c_3 =-2\cos
\vartheta_1-2\cos\vartheta_2\\
4-2 c_1-2 c_3+c_1c_3=4\cos\vartheta_1\cos\vartheta_2\end{array}\rt..
\eea

The above relation and the relations (\ref{Relation-t-tau}) between
$\bar\Lambda_m(u)$ and $\Lambda_m(u)$ lead to that the asymptotic
behaviors (\ref{asym-4-1})-(\ref{asym-4-3}) of the eigenvalues
$\bar\Lambda_m(u)$ are automatically satisfied. One can easily show
that the ansatz (\ref{4t1-103})-(\ref{4t3-103}) also make the very
functional relations (\ref{Eig-funtion-tau}) fulfilled. The regular
property of $\Lambda(u)$ leads to the associated BAEs \bea
&&1+\frac{\lambda_l^{(1)}} {\lambda_l^{(1)}+\eta}
\frac{K^{(2)}({\lambda_l^{(1)}})d(\lambda^{(1)}_l)}{K^{(1)}({\lambda_l^{(1)}})a(\lambda^{(1)}_l)}
\frac{Q^{(1)}(\lambda^{(1)}_l+\eta)Q^{(2)}(\lambda^{(1)}_l-\eta)}{Q^{(1)}(\lambda^{(1)}_l-\eta)Q^{(2)}(\lambda^{(1)}_l)}\no\\
&&\quad\quad\quad\quad+\frac{\lambda^{(1)}_l(\lambda^{(1)}_l+\frac{\eta}{2})d(\lambda^{(1)}_l)}{K^{(1)}(\lambda^{(1)}_l)}
\frac{F_1(\lambda^{(1)}_l)}{Q^{(1)}(\lambda^{(1)}_l-\eta)}=0,\, l=1,\ldots, L_1,\label{BAE-Open-4-1}\\[6pt]
&&\frac{\lambda_l^{(2)}+\frac{\eta}{2}}
{\lambda_l^{(2)}+\frac{3}{2}\eta}
\frac{K^{(3)}({\lambda_l^{(2)}})}{K^{(2)}({\lambda_l^{(2)}})}
\frac{Q^{(1)}(\lambda^{(2)}_l)Q^{(2)}(\lambda^{(2)}_l+\eta)Q^{(3)}(\lambda^{(2)}_l-\eta)}
{Q^{(1)}(\lambda^{(2)}_l+\eta)Q^{(2)}(\lambda^{(2)}_l-\eta)Q^{(3)}(\lambda^{(1)}_l)}=-1,\, l=1,\ldots,L_2,\label{BAE-Open-4-2}\\[6pt]
&&1+\frac{(\lambda^{(3)}_l+\eta)(\lambda^{(3)}_l+\frac{3}{2}\eta)a(\lambda^{(3)}_l)}{K^{(3)}(\lambda^{(3)}_l)}
\frac{Q^{(2)}(\lambda^{(3)}_l)F_3(\lambda^{(3)}_l)}{Q^{(2)}(\lambda^{(3)}_l+\eta)Q^{(3)}(\lambda^{(3)}_l-\eta)}\no\\
&&\quad\quad\quad\quad+\frac{\lambda_l^{(3)}+\eta}
{\lambda_l^{(3)}+2\eta}
\frac{K^{(4)}({\lambda_l^{(3)}})}{K^{(3)}({\lambda_l^{(3)}})}
\frac{Q^{(2)}(\lambda^{(3)}_l)Q^{(3)}(\lambda^{(3)}_l+\eta)}{Q^{(2)}(\lambda^{(3)}_l+\eta)Q^{(3)}(\lambda^{(3)}_l-\eta)}=0,
\, l=1,\ldots, L_3.\label{BAE-Open-4-3} \eea
It can be shown that the BAEs
(\ref{BAE-Open-4-1})-(\ref{BAE-Open-4-3}) also guarantee the
regularity of the ansatz $\Lambda_2(u)$ and $\Lambda_3(u)$ given by
(\ref{4t2-103}) and (\ref{4t3-103}), respectively. Moreover, the
ansatz (\ref{4t1-103})-(\ref{4t3-103}) indeed satisfy the relations
(\ref{t-4-1})-(\ref{t-4-12}). Finally, we conclude that the ansatz
$\Lambda_m(u)$ given by (\ref{4t1-103})-(\ref{4t3-103}) are the
eigenvalues of the transfer matrices $t_m(u)$ of the
$su(4)$-invariant open spin chain with the most general non-diagonal
boundary terms.

The eigenvalue of the Hamiltonian of the $su(4)$-invariant open chain with generic non-diagonal boundary terms is given by
\begin{eqnarray}
E= \sum_{l=1}^{L_1}
\frac{2\eta^2}{\lambda_l^{(1)}(\lambda_l^{(1)}+\eta)}
+2(N-1)+\eta\frac{[K^{(1)}(u)]^\prime}{K^{(1)}(u)}\mid_{u\rightarrow
0}+\frac 12,\no
\end{eqnarray}
where the parameters $\{\lambda^{(1)}_l\}$ are the roots of the BAEs
(\ref{BAE-Open-4-1})-(\ref{BAE-Open-4-3}) in the homogeneous limit
$\theta_j=0$.

\section{Exact solution of $su(n)$-invariant spin chain with general open boundaries}
\setcounter{equation}{0}

\subsection{Nested T-Q ansatz and Bethe ansatz equations}

Here  we present the result for  the $su(n)$-invariant quantum spin chain with
general open boundary conditions.
The functions  $z_m(u)$ read
\begin{eqnarray}
&&z_{m}(u)
=\frac{2u(2u+n\eta)}{(2u+(m-1)\eta)(2u+m\eta)}K^{(m)}(u)Q^{(0)}(u)
\frac{Q^{(m-1)}(u+\eta)Q^{(m)}(u-\eta)}{Q^{(m-1)}(u)Q^{(m)}(u)},
\label{2z2-11}\\
&&\quad\quad m=1, \ldots, n,\no
\end{eqnarray}
where $\{K^{(l)}(u)|l=1,\ldots,n\}$ satisfy
(\ref{keyk2})-(\ref{keyk1}). In principle, $K^{(l)}(u)$ could be any
decomposition of (\ref{keyk2}). For simplicity, we
parameterize them satisfying the property
\begin{eqnarray}
K^{(l)}(u) = K^{(l+1)}(-u-l\eta), \quad l=1, \cdots, n-1.
\end{eqnarray}
The function $x_m(u)$ is
\begin{eqnarray}
\lt\{\begin{array}{l}x_{2l-1}(u)= u(u+\frac n 2\eta)a(u)d(u)
\frac{F_{2l-1}(u)}{Q^{(2l-1)}(u)}\\
x_{2l}(u)=0\end{array}\rt.,
\end{eqnarray}
and $l=1, 2, \ldots, \frac{n}{2}$ if $n$ is even, $l=1, 2, \ldots,
\frac{n-1}{2}$ and $\tilde z_{n}(u)=z_{n}(u)$ if $n$ is odd.
The functions $\{F_{2l-1}(u)\}$ are given by
\begin{eqnarray}
&&F_{1}(u)=f_{1}(u)Q^{(2)}(-u-\eta),
\no \\
&&F_{2l-1}(u)=f_{2l-1}(u)Q^{(2l-2)}(-u-(2l-1)\eta)Q^{(2l)}(-u-(2l-1)\eta)a(-u-(2l-1)\eta),
\no
\end{eqnarray}
where $l=2, \ldots, \frac{n}{2}$ if $n$ is even and $l=2, \ldots,
\frac{n-1}{2}$ if $n$ is odd, and \bea
f_{2l-1}(u)=c_{2l-1}\prod_{k=1}^{n-1} (u+ \frac k 2 \eta)
(u+(2l-1)\eta- \frac{k}{2}\eta), \quad l=1, 2, \cdots . \eea The
functions $f_{2l-1}$ has the crossing symmetry \bea
f_{2l-1}(u)=f_{2l-1}(-u-(2l-1)\eta). \eea Here the parameters
$\{c_{2l-1}\}$ are  determined, with helps of the asymptotic
behaviors of the eigenvalues of the transfer matrices, by the
following relations
\begin{eqnarray}
&& \sum_{1\leq i_{1} < i_2 < \ldots < i_m\leq n } \tilde c_{i_1}
\tilde c_{i_2} \ldots \tilde  c_{i_m}
+\sum_{k=1}^{m_1}\sum_{l=k}^{m_2} \sum_{1\leq i_1 < i_2< \ldots <
i_{2k-2}\leq 2l-2} \tilde c_{i_1}\tilde c_{i_2} \ldots \tilde
c_{i_{2k-2}} \tilde c_{2l-1}
\no \\
&&  \times  \sum_{2l+1\leq i_{2k+1}<i_{2k+2}<\ldots <i_m\leq n}
\tilde c_{i_{2k+1}}\tilde c_{i_{2k+2}} \ldots \tilde
c_{i_{m}}+\sum_{k=2}^{m_3}\sum_{l=k}^{m_4} \sum_{1\leq i_1 < i_2<
\ldots < i_{2k-3}\leq 2l-2} \tilde c_{i_1} \no
\\
&& \times \tilde c_{i_2} \ldots \tilde c_{i_{2k-3}} \tilde c_{2l-1}
\sum_{2l+1\leq i_{2k}<i_{2k+1}<\ldots <i_m\leq n}
\tilde c_{i_{2k}}\tilde c_{i_{2k+1}} \ldots \tilde c_{i_{m}} \no \\
&&=(-1)^m  \sum_{1 \leq i_1<i_{2}<\ldots <i_{m} \leq n}
\lambda_{i_1}\lambda_{i_2} \ldots \lambda_{i_m},\label{r1-n-1}
\end{eqnarray}
where
\bea
\tilde c_i=\lt\{\begin{array}{lll}-1+ c_{2l-1},&i=2l-1,\\
-1,&i=2l, \\
-1,&i=n, \end{array}\rt.\no \eea and
\begin{eqnarray}
&& (i). \;\;\;\; m_1=m_3=\frac{m}{2}, \quad
m_2=m_4=\frac{n-m-2k}{2}, \quad \text{if $m$ is
even and $n$ is even}; \label{1217-11}\\
&& (ii). \;\;\; m_1=\frac{m-1}{2},\quad  m_2=m_4=\frac{n-m-2k-1}{2},
\quad
m_3=\frac{m+1}{2}, \no \\
&&\qquad\qquad \text{ if $m$ is odd and $n$ is even}, \\
&& (iii). \;\; m_1=m_3=\frac{m}{2}, \quad
m_2=m_4=\frac{n-m-2k-1}{2}, \no \\
&&\qquad \qquad \text{ if $m$ is
even and $n$ is odd},\\
&& (iv). \;\;\; m_1=\frac{m-1}{2},\quad  m_2=\frac{n-m-2k}{2}, \quad
m_3=\frac{m+1}{2}, \quad
m_4=\frac{n-m-2k-2}{2}, \no \\
&&\qquad \qquad \text{ if $m$ is odd and $n$ is odd}.\label{1217-41}
\end{eqnarray}
Then the nested T-Q ansatz of
the eigenvalues $\Lambda(u)$ of  the transfer matrix $t(u)$ is
\begin{eqnarray}
\Lambda(u)= \sum_{i_1=1}^{n} \tilde z_{i_1}(u).
\label{newanatz1-1-21-2}
\end{eqnarray}
All the eigenvalues $\Lambda_m(u)$ of the fused transfer matrix $t_m(u)$ are given by
\begin{eqnarray}
&&\Lambda_m(u)=
\prod_{l=1}^{m-1}\prod_{k=1}^l\rho_2(2u-k\eta-l\eta+\eta)\no
\\ &&\qquad \times \lt\{\sum_{1\leq i_{1} < i_2 < \ldots < i_m\leq n } \tilde
z_{i_1}(u) \tilde z_{i_2}(u-\eta) \ldots \tilde z_{i_m}(u-(m-1)\eta)\rt. \no \\
&&\qquad -\sum_{k=1}^{m_1}\sum_{l=k}^{m_2} \sum_{1\leq i_1 < i_2<
\ldots < i_{2k-2}\leq 2l-2}  \tilde z_{i_1}(u)
 \tilde z_{i_2}(u-\eta) \ldots  \tilde z_{i_{2k-2}}(u-(2k-3)\eta) \no \\
&&\qquad \times f_{2l-1}(u-(2k-2)\eta)  \tilde z_{2l}(u-(2k-1)\eta) \sum_{2l+1\leq i_{2k+1}<i_{2k+2}<\ldots <i_m\leq n} \no \\
&&\qquad \times \tilde z_{i_{2k+1}}(u-2k\eta)\tilde
z_{i_{2k+2}}(u-(2k+1)\eta) \ldots \tilde z_{i_{m}}(u-(m-1)\eta) \no
\\
&&\qquad -\sum_{k=2}^{m_3}\sum_{l=k}^{m_4} \sum_{1\leq i_1 < i_2<
\ldots < i_{2k-3}\leq 2l-2}  \tilde z_{i_1}(u)
 \tilde z_{i_2}(u-\eta) \ldots  \tilde z_{i_{2k-3}}(u-(2k-4)\eta) \no \\
&&\qquad \times f_{2l-1}(u-(2k-3)\eta)  \tilde z_{2l}(u-(2k-2)\eta)\sum_{2l+1\leq i_{2k}<i_{2k+1}<\ldots <i_m\leq n} \no \\
&&\qquad \times \lt.\tilde z_{i_{2k}}(u-(2k-1)\eta)\tilde
z_{i_{2k+1}}(u-2k\eta) \ldots \tilde z_{i_{m}}(u-(m-1)\eta)\rt\},
\label{newanatztiltm-2-1107-even-n-enen-m}
\end{eqnarray}
where the $m_{1,\cdots,4}$ is the same as that in Eqs.
(\ref{1217-11})-(\ref{1217-41}). The parameters
$\{\lambda^{(r)}_l\}$ satisfy the associated Bethe ansatz equations
\begin{eqnarray}
&&K^{(1)}(\lambda^{(1)}_j)a(\lambda^{(1)}_j)Q^{(1)}(\lambda^{(1)}_j-\eta)
+\frac{\lambda^{(1)}_j}{\lambda^{(1)}_j+\eta}K^{(2)}(\lambda^{(1)}_j)d(\lambda^{(1)}_j)Q^{(1)}(\lambda^{(1)}_j+\eta)
\frac{Q^{(2)}(\lambda_j^{(1)}-\eta)}{Q^{(2)}(\lambda_j^{(1)})}\no \\
&&\qquad +
\lambda^{(1)}_j(\lambda^{(1)}_j+\frac{\eta}{2})a(\lambda^{(1)}_j)d(\lambda^{(1)}_j)
F_1(\lambda^{(1)}_j)=0, \quad j=1, \ldots, L_1.
\label{sunbaet1-1107-1}
\end{eqnarray}
\begin{eqnarray}
&&\frac{2\lambda_k^{(2l)}+(2l+1)\eta}{2\lambda_k^{(2l)}+(2l-1)\eta}
\frac{K^{(2l)}(\lambda_k^{(2l)})}{ K^{(2l+1)}(\lambda_k^{(2l)})}
\frac{Q^{(2l-1)}(\lambda_k^{(2l)}+\eta)Q^{(2l+1)}(\lambda_k^{(2l)})}
{Q^{(2l-1)}(\lambda_k^{(2l)})Q^{(2l+1)}(\lambda_k^{(2l)}-\eta)}
=-\frac{Q^{(2l)}(\lambda_k^{(2l)}+\eta)}{Q^{(2l)}(\lambda_k^{(2l)}-\eta)},
\no \\
&&\qquad \qquad  k=1,\ldots, L_{2l}, \label{sunbaet1-1107-2}
\end{eqnarray}
\begin{eqnarray}
&&K^{(2s+1)}(\lambda^{(2s+1)}_j)Q^{(2s+1)}(\lambda^{(2s+1)}_j-\eta)
+\frac{\lambda^{(2s+1)}_j+s\eta}{\lambda^{(2s+1)}_j+(s+1)\eta}
K^{(2s+2)}(\lambda^{(2s+1)}_j) \no
\\
&&\qquad \times Q^{(2s+1)}(\lambda^{(2s+1)}_j+\eta)
\frac{Q^{(2s)}(\lambda_j^{(2s+1)})Q^{(2s+2)}(\lambda_j^{(2s+1)}-\eta)}
{Q^{(2s)}(\lambda_j^{(2s+1)}+\eta)Q^{(2s)}(\lambda_j^{(2s+1)})}+
(\lambda^{(2s+1)}_j+s\eta) \no \\
&&\qquad \times
(\lambda^{(2s+1)}_j+\frac{2s+1}{2}\eta)a(\lambda^{(2s+1)}_j)
\frac{Q^{(2s)}(\lambda_j^{(2s+1)})}{Q^{(2s)}(\lambda_j^{(2s+1)}+\eta)}
F_{2s+1}(\lambda^{(2s+1)}_j)=0, \label{sunbaet1-1107-3} \\
&& \qquad \qquad j=1, \ldots, L_{2s+1}, \no
\end{eqnarray}
where $l=s=1, \ldots, \frac{n}{2}-1$ if $n$ is even, $l=1, \ldots,
\frac{n-1}{2}$ and $s=1, \ldots, \frac{n-1}{2}-1$ if $n$ is odd.

The eigenvalue of the Hamiltonian (\ref{oh}) is
\begin{eqnarray}
E= \sum_{l=1}^{L_1}
\frac{2\eta^2}{\lambda_l^{(1)}(\lambda_l^{(1)}+\eta)}
+2(N-1)+\eta\frac{[K^{(1)}(u)]^\prime}{K^{(1)}(u)}\mid_{u\rightarrow
0}+\frac 2 n,
\end{eqnarray}
where the parameters $\{\lambda^{(1)}_l\}$ are the roots of the BAEs
(\ref{sunbaet1-1107-1})-(\ref{sunbaet1-1107-3}) with $\theta_j=0$.

\subsection{Reduction to the diagonal boundary terms}

When two $K$-matrices $K^+(u)$ and $K^-(u)$ are both diagonal
matrices,  or they can be diagonalized  simultaneously by some gauge
transformation, all the parameters $c_{2l-1}$ vanish, leading to
$F_{2l-1}(u)=0$. The nested T-Q ansatz of the $\Lambda(u)$ in this
case becomes
\begin{eqnarray}
&&\Lambda(u)=\frac{2u+n\eta}{2u+\eta} K^{(1)}(u)Q^{(0)}(u+\eta)
\frac{Q^{(1)}(u-\eta)}{Q^{(1)}(u)}
\nonumber \\
&& \quad+
\frac{2u(2u+n\eta)}{(2u+\eta)(2u+2\eta)}K^{(2)}(u)Q^{(0)}(u)\frac{Q^{(1)}(u+\eta)Q^{(2)}(u-\eta)}{Q^{(1)}(u)Q^{(2)}(u)}
+ \ldots
\nonumber \\
&&\quad+\frac{2u(2u+n\eta)}{(2u+(n-2)\eta)(2u+(n-1)\eta)}K^{(n-1)}(u)
Q^{(0)}(u)\frac{Q^{(n-2)}(u+\eta)Q^{(n-1)}(u-\eta)}{Q^{(n-2)}(u)Q^{(n-1)}(u)} \no \\
&&\quad + \frac{2u}{2u+(n-1)\eta} Q^{(0)}(u)K^{(n)}(u)
\frac{Q^{(n-1)}(u+\eta)}{Q^{(n-1)}(u)}.\label{open-anatz1-1-1}
\end{eqnarray}
For example, $K^{(l)}(u)$ can be parameterized as
\begin{eqnarray}
&&K^{(l)}(u)=K^{(l)}_+(u)\,K^{(l)}_-(u),\\
&& K_+^{(l)}(u)=\bar \xi -u\bar \varepsilon_l-\frac{1}{2}\eta(l
\bar\varepsilon_l+\bar\varepsilon_{l+1}+\ldots+\bar\varepsilon_n),
\label{10-16-4}
\\
&& K_-^{(l)}(u)=\xi +u\varepsilon_l
+\frac12\eta[(l-1)\varepsilon_l-\varepsilon_1-\varepsilon_2-\ldots-\varepsilon_{l-1}],\quad
l=1, \ldots, n, \label{10-16-5}
\end{eqnarray}
where $\varepsilon_l=\pm 1$ and $\bar \varepsilon_l=\pm 1$. The regularity of $\Lambda(u)$
leads to the associated BAEs
\begin{eqnarray}
&&\frac{Q^{(1)}(\lambda_j^{(1)}-\eta)}{Q^{(1)}(\lambda_j^{(1)}+\eta)}
=-\frac{\lambda_j^{(1)}}{\lambda_j^{(1)}+\eta}
\frac{K^{(2)}(\lambda_j^{(1)})}{K^{(1)}(\lambda_j^{(1)}) }
\frac{d(\lambda_j^{(1)})}{a(\lambda_j^{(1)})}
\frac{Q^{(2)}(\lambda_j^{(1)}-\eta)}{Q^{(2)}(\lambda_j^{(1)})},
\quad j=1,\ldots, L_1, \label{BAE-open-diagonal-1} \\[6pt]
&&\frac{
Q^{(r)}(\lambda_k^{(r)}-\eta)}{Q^{(r)}(\lambda_k^{(r)}+\eta)} =-
\frac{K^{(r+1)}(\lambda_k^{(r)})}{K^{(r)}(\lambda_k^{(r)})}
\frac{2\lambda_k^{(r)}+(r-1)\eta}{2\lambda_k^{(r)}+(r+1)\eta}
\frac{Q^{(r-1)}(\lambda_k^{(r)})Q^{(r+1)}(\lambda_k^{(r)}-\eta)}
{Q^{(r-1)}(\lambda_k^{(r)}+\eta) Q^{(r+1)}(\lambda_k^{(r)})},
\label{BAE-open-diagonal-2} \\[6pt]
&& \qquad \qquad k=1,\ldots, L_r, \quad r=2,\ldots, n-1. \no
\end{eqnarray}
Then the resulting expression (\ref{open-anatz1-1-1}) of the
eigenvalue $\Lambda(u)$ of the transfer matrix $t(u)$ and the BAEs
(\ref{BAE-open-diagonal-1})-(\ref{BAE-open-diagonal-2}) recover
those obtained by the other Bethe ansatz methods
\cite{Yan04,Mel050,Mel051,Mel052,Mel053,Mel054,Mel055,Mel057}. In
this case $\Lambda(u)$ can be expressed in terms of the functions
$\{z_l(u)\}$ given by (\ref{2z2-11})
\begin{eqnarray}
\Lambda(u)=\sum_{l=1}^{n} z_{l}(u). \label{openanatz1-1}
\end{eqnarray}
Moreover, the eigenvalues $\Lambda_m(u)$ of the fused
transfer matrices $t_m(u)$ can be  constructed by the functions $\{z_l(u)\}$ as
\begin{eqnarray}
\Lambda_m(u)&=&\prod_{l=1}^{m-1}\prod_{k=1}^l\rho_2(2u-k\eta-l\eta+\eta)\no\\
&&\quad\quad \times \sum_{1\leq i_1<i_2<\ldots < i_m \leq n}
z_{i_1}(u)z_{i_2}(u-\eta) \ldots z_{i_m}(u-(m-1)\eta).
\label{openanatz1-2}
\end{eqnarray}
The BAEs  (\ref{BAE-open-diagonal-1})-(\ref{BAE-open-diagonal-2})
guarantee the regularities of the expressions (\ref{openanatz1-2}) of  the eigenvalues $\Lambda_m(u)$ for the higher fused transfer matrices.


\section{Conclusions}
In this paper, we propose the nested off-diagonal Bethe ansatz
method for solving the multi-component integrable models with
generic integrable boundaries, a generalization of the method
proposed in \cite{cao1} (related to $su(2)$ algebra) for integrable
models associated with higher rank algebras.  In the method some
functional relations (for the $su(n)$ case such as  (\ref{id1}) for
the closed chain or (\ref{openfun1}) for the open chain) among the
antisymmetric fused transfer matrices play a very important role.
Taking the $su(n)$-invariant spin chain model with both periodic and
non-diagonal boundaries as examples, we elucidate how the method
works for constructing the Bethe ansatz solutions of the model. For
the $su(n)$-invariant closed chain, our results (\ref{anatz1-1-1})
and (\ref{sunbea2210}) recover those obtained via other BA methods
\cite{3-sun,3-sun1} but with a greatly simplified process. For the
open boundary case specified by the most general $K$-matrices
(\ref{km-}) and (\ref{km+++}), the very functional relations
(\ref{openfun1}) are derived only via some properties of the
$R$-matrix and $K$-matrices. Based on these relations, the
asymptotic behaviors (\ref{Asymptotic-open}) and the values of the
eigenvalue functions at $\sum_{m=1}^{n-1}2m$ special points (such as
(\ref{t1-1})-(\ref{t2-4}) or (\ref{t-4-1})-(\ref{t-4-12})), we
obtain the eigenvalues of the transfer matrix. When the $K$-matrices
are both diagonal ones, our results can be reduced to those obtained
by the conventional Bethe ansatz methods. Therefore, our method
provides an unified procedure for approaching the integrable models
both with and without $U(1)$ symmetry. We remark that this method
might also be applied to other quantum integrable models defined in
different algebras.

\section*{Acknowledgments}

The financial support from  the National Natural Science Foundation
of China (Grant Nos. 11174335, 11031005, 11375141,
11374334), the National Program for Basic Research of MOST (973
project under grant No.2011CB921700), the State Education
Ministry of China (Grant No. 20116101110017) and BCMIIS are
gratefully acknowledged. Two of the authors (W.-L. Yang and K. Shi)
would like to thank IoP, CAS for the hospitality and they enjoyed
during their visit there. W.-L. Yang also would like to thank KITPC
for his  hospitality where some parts of work have been finished
during the visiting.

\section*{Appendix A: Proofs of the operator identities}
\setcounter{equation}{0}
\renewcommand{\theequation}{A.\arabic{equation}}

In this appendix, we give the detailed proof of the following
identities which are crucial to obtain the functional relations
(\ref{id1}) and (\ref{openfun1}): \bea T_1(\theta_j)T_{\langle
2,3,\ldots,m\rangle}(\theta_j-\eta)
&=& P^{(-)}_{1,2,\ldots, m}T_1(\theta_j)T_2(\theta_j-\eta)\ldots T_m(\theta_j-(m-1)\eta)\,P^{(-)}_{2,\ldots, m},\label{A.1}\\
{\cal{J}}_{1,\ldots,m}(\pm\theta_j)&=&P^{(-)}_{1,2,\ldots, m}{\cal{J}}_{1,\ldots,m}(\pm\theta_j).\label{A.2}
\eea
The exchange relation (\ref{RTT}) of the one row monodromy matrix $T(u)$ implies
\bea
R_{\bar{2}\bar{1}}(-\eta)T_{\bar{2}}(u-\eta)T_{\bar{1}}(u)
=T_{\bar{1}}(u)T_{\bar{2}}(u-\eta)R_{\bar{2}\bar{1}}(-\eta).\no
\eea The above relation and the fusion condition (\ref{Fusion}) allow one  to derive the following identity
\bea
P^{(-)}_{\bar{1},\bar{2}}\,T_{\bar{1}}(u)T_{\bar{2}}(u-\eta)R_{\bar{1}\bar{2}}(-\eta)
=T_{\bar{1}}(u)T_{\bar{2}}(u-\eta)R_{\bar{1}\bar{2}}(-\eta).\label{A.3}
\eea Let us evaluate the product of the operators $T_{\bar{1}}(\theta_j)$ and
$T_{\bar{2}}(\theta_j\hspace{-0.02truecm}-\hspace{-0.02truecm}\eta)$
\bea
T_{\bar{1}}(\theta_j)T_{\bar{2}}(\theta_j\hspace{-0.02truecm}-\hspace{-0.02truecm}\eta)
\hspace{-0.6truecm}&=&\hspace{-0.4truecm}R_{\bar{1} N}(\theta_j\hspace{-0.02truecm}-\hspace{-0.02truecm}\theta_N)\ldots
R_{\bar{1}\,j+1}(\theta_j\hspace{-0.02truecm}-\hspace{-0.02truecm}\theta_{j+1})
R_{\bar{1} j}(0)
R_{\bar{1}\,j-1}(\theta_j\hspace{-0.02truecm}-\hspace{-0.02truecm}\theta_{j-1})
\ldots R_{\bar{1}\,1}(\theta_j\hspace{-0.02truecm}-\hspace{-0.02truecm}\theta_{1})\no\\
&&\hspace{-0.2truecm}\times
R_{\bar{2} N}(\theta_j\hspace{-0.02truecm}-\hspace{-0.02truecm}\theta_N-\eta)\ldots
R_{\bar{2}\,j+1}(\theta_j\hspace{-0.02truecm}-\hspace{-0.02truecm}\theta_{j+1}-\eta)
R_{\bar{2} j}(-\eta)\no\\
&&\hspace{-0.2truecm}\times R_{\bar{2}\,j-1}(\theta_j\hspace{-0.02truecm}-\hspace{-0.02truecm}\theta_{j-1}-\eta)
\ldots R_{\bar{2}\,1}(\theta_j\hspace{-0.02truecm}-\hspace{-0.02truecm}\theta_{1}-\eta)\no\\
\hspace{-0.6truecm}&=&\hspace{-0.4truecm}R_{j\,j-1}(\theta_j\hspace{-0.02truecm}-\hspace{-0.02truecm}\theta_{j-1})
\ldots R_{j\,1}(\theta_j\hspace{-0.02truecm}-\hspace{-0.02truecm}\theta_{1})\no\\
&&\hspace{-0.2truecm}\times R_{\bar{1} N}(\theta_j\hspace{-0.02truecm}-\hspace{-0.02truecm}\theta_N)\ldots
R_{\bar{1} j+1}(\theta_j\hspace{-0.02truecm}-\hspace{-0.02truecm}\theta_{j+1})\no\\
&&\hspace{-0.2truecm}\times R_{\bar{2} N}(\theta_j\hspace{-0.02truecm}-\hspace{-0.02truecm}\theta_N-\eta)\ldots
R_{\bar{2} j+1}(\theta_j\hspace{-0.02truecm}-\hspace{-0.02truecm}\theta_{j+1}-\eta)
R_{\bar{2}\bar{1}}(-\eta)\no\\
&&\hspace{-0.2truecm}\times R_{\bar{1}j}(0)
R_{\bar{2}\,j-1}(\theta_j\hspace{-0.02truecm}-\hspace{-0.02truecm}\theta_{j-1}\hspace{-0.02truecm}-\hspace{-0.02truecm}\eta)
\ldots R_{\bar{2}\,1}(\theta_j\hspace{-0.02truecm}-\hspace{-0.02truecm}\theta_{1}\hspace{-0.02truecm}-\hspace{-0.02truecm}\eta)\no\\
\hspace{-0.6truecm}&\stackrel{(\ref{A.3})}{=}&\hspace{-0.4truecm}
R_{j\,j-1}(\theta_j\hspace{-0.02truecm}-\hspace{-0.02truecm}\theta_{j-1})
\ldots R_{j\,1}(\theta_j\hspace{-0.02truecm}-\hspace{-0.02truecm}\theta_{1})\,P^{(-)}_{\bar{1},\bar{2}}\no\\
&&\hspace{-0.2truecm}\times R_{\bar{1} N}(\theta_j\hspace{-0.02truecm}-\hspace{-0.02truecm}\theta_N)\ldots
R_{\bar{1} j+1}(\theta_j\hspace{-0.02truecm}-\hspace{-0.02truecm}\theta_{j+1})\no\\
&&\hspace{-0.2truecm}\times
R_{\bar{2} N}(\theta_j\hspace{-0.02truecm}-\hspace{-0.02truecm}\theta_N-\eta)\ldots
R_{\bar{2} j+1}(\theta_j\hspace{-0.02truecm}-\hspace{-0.02truecm}\theta_{j+1}-\eta)
R_{\bar{2}\bar{1}}(-\eta)\no\\
&&\hspace{-0.2truecm}\times R_{\bar{1}j}(0)
R_{\bar{2}\,j-1}(\theta_j\hspace{-0.02truecm}-\hspace{-0.02truecm}\theta_{j-1}\hspace{-0.02truecm}-\hspace{-0.02truecm}\eta)
\ldots R_{\bar{2}\,1}(\theta_j\hspace{-0.02truecm}-\hspace{-0.02truecm}\theta_{1}\hspace{-0.02truecm}-\hspace{-0.02truecm}\eta)\no\\
\hspace{-0.6truecm}&=&\hspace{-0.4truecm}P^{(-)}_{\bar{1},\bar{2}}\,
T_{\bar{1}}(\theta_j)T_{\bar{2}}(\theta_j\hspace{-0.02truecm}-\hspace{-0.02truecm}\eta),\no
\eea   namely, we have
\bea
 T_{1}(\theta_j)T_{2}(\theta_j\hspace{-0.02truecm}-\hspace{-0.02truecm}\eta)=
 P^{(-)}_{1,2}\,
T_{1}(\theta_j)T_{2}(\theta_j\hspace{-0.02truecm}-\hspace{-0.02truecm}\eta),\quad j=1,\ldots,N.\label{A.4}
\eea Similarly, we have
\bea
\hat{T}_{1}(-\theta_j)\hat{T}_{2}(-\theta_j\hspace{-0.02truecm}-\hspace{-0.02truecm}\eta)=
 P^{(-)}_{1,2}\,
\hat{T}_{1}(-\theta_j)\hat{T}_{2}(-\theta_j\hspace{-0.02truecm}-\hspace{-0.02truecm}\eta),\quad j=1,\ldots,N.\label{A.5}
\eea
Due to the fact that $R_{12}(-\eta)$ is proportional to the antisymmetric projector (\ref{Fusion}), the relation (\ref{A.3})
also implies
\bea
T_{\langle 1,2\rangle }(u)=P^{(-)}_{1,2}T_1(u)T_2(u-\eta)P^{(-)}_{1,2}=T_1(u)T_2(u-\eta)P^{(-)}_{1,2}.
\eea   Using similar method to derive the above relation and following the procedure \cite{Kar79}, we can derive the following relations
\bea
 T_{\langle 1,2,\ldots,m\rangle }(u)&=&P^{(-)}_{1,2,\ldots, m}T_1(u)T_2(u-\eta)\ldots T_m(u-(m-1)\eta)P^{(-)}_{1,2,\ldots, m}\no\\
 &=& T_1(u)T_2(u-\eta)\ldots T_m(u-(m-1)\eta)P^{(-)}_{1,2,\ldots, m}. \label{A.7}
\eea Combining the above relation with (\ref{A.4}), we can show that
\bea
P_{l,l+1}\,T_1(\theta_j)T_{\langle 2,3,\ldots,m\rangle }(\theta_j-\eta)=-T_1(\theta_j)T_{\langle 2,3,\ldots,m\rangle }(\theta_j-\eta),\quad l=1,\ldots,m-1.
\eea Then we can conclude that $T_1(\theta_j)T_{\langle 2,3,\ldots,m\rangle }(\theta_j-\eta)$ satisfy the relation (\ref{A.1}).

With the similar method used to prove (\ref{A.4}) and the reflection
equation (\ref{RE-V}), we can obtain the following relations: \bea
{\cal{J}}_1(\pm\theta_j)R_{21}(\pm
2\theta_j-\eta){\cal{J}}_2(\pm\theta_j-\eta)&=&
P^{(-)}_{1,2}{\cal{J}}_1(\pm\theta_j)R_{21}(\pm 2\theta_j-\eta){\cal{J}}_2(\pm\theta_j-\eta),\no\\
j&=&1,\ldots,N.\label{A.9}\\
{\cal{J}}_{\langle 1,\ldots,m\rangle }(u)&=&{\cal{J}}_{1,\ldots,m}(u)\,P^{(-)}_{1,\ldots, m},\,m=1,\ldots n,\label{A.10}\\
K^{+}_{\langle 1,\ldots,m\rangle
}(u)&=&K^{+}_{1,\ldots,m}(u)\,P^{(-)}_{1,\ldots, m},\,m=1,\ldots
n.\label{A.11} \eea Using the relations (\ref{A.9}) and
(\ref{A.10}), we can  derive that \bea
P_{l,l+1}\,{\cal{J}}_{1,\ldots,m}(\pm\theta_j)=-{\cal{J}}_{1,\ldots,m}(\pm\theta_j),\quad
l=1,\ldots,m-1. \eea (\ref{A.2}) is a consequence of the above
relations. Hence we complete the proof of (\ref{A.2}).

\end{document}